\documentclass[preprint,amsmath,amssymb,aps]{revtex4-2}
\usepackage{graphicx}
\usepackage{tabularx}
\usepackage{bm}
\usepackage{txfonts}
\usepackage{hyperref}
\usepackage{caption}
\usepackage{multirow}
\usepackage{subcaption}
\usepackage{soul}
\date{\today}
\usepackage{xcolor}
\newcommand{\be}{\begin{eqnarray}}
\newcommand{\ee}{\end{eqnarray}}

\newcommand{\bfk}{{\bf k}_{\perp}}

\usepackage{xcolor}
\begin{document}
\title{Spectroscopy of excited quarkonium states in the light-front quark model}

\author{Ritwik Acharyya}
\email{ritwikacharyya2001@gmail.com}
\affiliation{Computational High Energy Physics Lab, Department of Physics, Dr. B R Ambedkar National
	Institute of Technology Jalandhar, Punjab, India, 144008}

\author{Satyajit Puhan}
\email{puhansatyajit@gmail.com}
\affiliation{Computational High Energy Physics Lab, Department of Physics, Dr. B R Ambedkar National
	Institute of Technology Jalandhar, Punjab, India, 144008}

\author{Narinder Kumar}
\email{narinderhep@gmail.com}
\affiliation{Computational Theoretical High Energy Physics Lab, Department of Physics, Doaba College, Jalandhar 144004, India}
\affiliation{Computational High Energy Physics Lab, Department of Physics, Dr. B R Ambedkar National
	Institute of Technology Jalandhar, Punjab, India, 144008}

\author{Harleen Dahiya}
\email{dahiyah@nitj.ac.in}
\affiliation{Computational High Energy Physics Lab, Department of Physics, Dr. B R Ambedkar National
	Institute of Technology Jalandhar, Punjab, India, 144008}

\date{\today}%
\begin{abstract}

We have investigated the ground state ($1S$), radially excited states ($2S$) and ($3S$) along with the orbitally excited state ($1P$) for the heavy charmonia ($c \bar c$) and bottomonia ($b \bar b$) mesons in the light-front quark model (LFQM). The light-front wave functions have been successful in explaining various physical properties of meson states in the past, especially for the $1S$ and $2S$ states. However, studies regarding the radially excited state $3S$ and orbitally excited state $1P$ have hardly been pursued before. In this study, we take up these two excited states and investigate the electromagnetic form factors (EMFFs), charge radii, decay constants, parton distribution functions (PDFs) and the distribution amplitudes (DAs) for the quarkonia system. For the sake of completeness, we have also included the study of the ground and the first excited states of quarkonia mesons. We have also illustrated the 3D wave functions for the radially excited states in order to study their nodal structures.  

 \vspace{0.1cm}
\end{abstract}
%
\maketitle
%
%
\section{Introduction\label{secintro}}

Quantum Chromodynamics (QCD) \cite{Marciano:1977su,Altarelli:1981ax} provides us an understanding of the intricacies of the strong interactions that occur in our universe. The subject of QCD explains how a parent hadron comprises the fundamental particles namely the quarks, gluons and sea quarks as well as the interactions that exist among each of them. In the recent years, understanding the multi-dimensional structure of the hadrons has been an interesting topic \cite{Angeles-Martinez:2015sea, Boer:1997nt,Boer:1999mm, Ralston:1979ys, Kotzinian:1994dv, Radici:2014bfa, Sivers:1989cc,Belitsky:2005qn,Diehl:2015uka, Chakrabarti:2005zm,Brodsky:2006ku,Rajan:2016tlg,Hagler:2003jw,Ji:2004gf,Conti:2009zza,Martin:1998sq,Gluck:1998xa,Gluck:1994uf,Mulders:1995dh,Garcon:2002jb,Belitsky:2005qn,Puhan:2023hio}. To visualize the partonic content and its behavior within a parent hadron, one employs different mathematical constructs such as the parton distribution functions (PDFs) \cite{Conti:2009zza,Martin:1998sq,Gluck:1998xa,Gluck:1994uf,Mulders:1995dh,Garcon:2002jb} which gives us a one dimensional (1-D) picture of the hadron and describe the likelihood of  finding a parton carrying the parent hadron's longitudinal momentum fraction ($x$). The PDFs, however,  do not carry any information about other degree of freedom of the parton. To obtain the two-dimensional (2-D) representation of the hadron, one can study the form factors (FFs) giving us information regarding the charge distribution inside a hadron. Further, one can achieve  deeper insights for spatial and transverse structure on the hadron by investigating their three dimensional (3-D) structure using the transverse-momentum dependent parton distributions (TMDs) \cite{Angeles-Martinez:2015sea, Boer:1997nt,Boer:1999mm, Ralston:1979ys, Kotzinian:1994dv, Radici:2014bfa, Sivers:1989cc,Puhan:2023hio} and the generalized parton distributions (GPDs) \cite{Garcon:2002jb,Belitsky:2005qn,Diehl:2015uka, Chakrabarti:2005zm,Brodsky:2006ku,Rajan:2016tlg,Hagler:2003jw,Ji:2004gf}. 

The distribution amplitude (DA) provides detailed information on the strength of the coupling between quark and anti-quark pairs within a hadron. It quantifies the probability amplitude for finding these constituent particles bound together, thereby indicating the effectiveness of their interaction in forming the hadronic structure. 
In quantum field theory, these non-perturbative and scale-dependent functions can be thought of as the closest relatives of quantum mechanical wave functions \cite{Serna:2020txe}. In the limit of zero transverse momentum, represented by the leading-twist DAs, the leading Fock-state contribution to its light-front wave function is considered. Since particle number is conserved in this frame, the light-front formulation of a wave function permits a probability interpretation of partons that is not easily available in an infinite-body field theory.

Light-front dynamics (LFD) plays a crucial role in understanding the theory of strong interactions. It is a dynamical theory where the dynamical variables correspond to the physical conditions on a front $x^{+} = 0$ \cite{Harindranath:1996hq}. In particular, because of its unique characteristics of rational energy-momentum dispersion relation, LFD is discovered to offer an efficient means of handling the relativistic effects. With a maximum number of seven kinematic (or interaction-independent) generators, it requires less dynamic effort to obtain QCD solutions that fully reflect Poincaré symmetries \cite{dirac:1949}. In this paper, we have adopted the light-front quark model (LFQM) \cite{Jaus:1989au, Jaus:1991cy,Puhan:2024ckp, Coester:2005cv,Puhan:2023ekt,Acharyya:2024enp} which describes a hadron's composition as a bound state between its active quark and anti-quark spectator. LFQM focuses on hadron's valence quarks: the fundamental elements responsible for the overall structure and behavior. Since both the light-front transverse and the longitudinal boost operators are kinematical, the light-front wave function is independent of the external momentum.

Even though LFQM has already been used to successfully explain the various characteristics of the ground state mesons \cite{Cheng:1997au,Hwang:2001th, Arndt:1999wx, Choi:1997iq}, the attributes of the radially excited states are still pending investigation given that, in contrast to the ground states, their nature is currently unknown and less understood. Radially excited states of hadrons in particular provide complementary information to the orbitally excited states, making them crucial to comprehending the strong interactions. The Particle Data Group (PDG) \cite{ParticleDataGroup:2020ssz} states that a few of them have not yet been verified, but they have been seen in the light and heavy quark sectors of hadrons. In the present work, we have investigated the wave functions of the ground state ($1S$), radially excited states ($2S$ and $3S$) and the orbitally excited state ($1P$) of the bottomonia ($b\bar b$) and charmonia ($c\bar c$) for spin-0 and spin-1 mesons within the LFQM regime. We have discussed various characteristics related to them such as the DAs, PDFs, decay constants,  electromagnetic form factors (EMFFs) and charge radii. It would be important to mention here that these meson attributes have been studied previously for the ground state \cite{Cheng:1997au,Braga:2015lck,Plant:1997jr} and the radially excited state $2S$ \cite{Chen:2021kby} but are yet to be investigated for the $3S$ and $1P$ states. We compare our results obtained between the pseudo-scalar and vector mesons for their various excited states. Therefore, the major goal of our work is to examine these features and assess how they differ from heavy quarkonia in the ground state. We believe that our results will help to obtain a better understanding of the excited states of mesons in future.

The paper is arranged as follows. In Sec. \ref{meth}, we have presented quantitatively the LFQM, which is the primary model we consider for this study. In Sec. \ref{prop}, we have presented our chosen input parameters and have also theorized various structural properties of the meson studied in this paper. Further, we have discussed the numerical results obtained for the various excited states of the heavy quarkonia meson systems. We have also illustrated plots showing the nature of  different wave functions that we have considered here. Finally, we have summarized our results in Sec. \ref{sum}.


 \section{Light-front quark model} \label{meth}

 The minimal hadron Fock-state wave function based on light-front quantization in the form of quark-antiquark is expressed as \cite{Puhan:2023ekt,Kaur:2020vkq}
\begin{eqnarray}
|M(P, J, J_{z})\rangle &=& \sum_{\lambda_q,\lambda_{\bar q}}\int
\frac{\mathrm{d} x \ \mathrm{d}^2
        \mathbf{k}_{\perp}}{\sqrt{x(1-x)}16\pi^3}
           \psi_{\lambda_q,\lambda_{\bar q}}^{J J_{z}}\left(x, \mathbf{k}_{\perp}\right)|x,\mathbf{k}_{\perp},
        \lambda_q,\lambda_{\bar q} \rangle
        ,
        \label{meson}
\end{eqnarray}
where $|M(P, J, J_{z})\rangle$ is the meson eigenstate with $P=(P^+,P^-,P_{\perp})$ being the four-vector total momentum of the hadron, $(J, J_{z})$ is the total angular momentum, $\textbf{k}=(k^+,k^-,\textbf{k}_{\perp})$ is the momentum of the active quark, $x=\frac{k^+}{P^+}$ is the longitudinal momentum fraction carried by the active quark and $\lambda_{q (\bar q)}$ is the helicity of quark (anti-quark).

The light-front wave function $\psi_{\lambda_q,\lambda_{\bar q}}^{J J_{z}}$ for pseudo-scalar and vector mesons is defined as follows
\begin{equation} \label{Twave}
\psi_{\lambda_q,\lambda_{\bar q}}^{J J_{z}}\left(x, \mathbf{k}_{\perp}\right)=\phi_{nS(nP)}\left(x, \mathbf{k}_{\perp}\right) \mathcal{X}_{\lambda_{q} \lambda_{\bar{q}}}^{J J_{z}}\left(x, \mathbf{k}_{\perp}\right),
\end{equation}
where $\phi_{nS(nP)}\left(x, \mathbf{k}_{\perp}\right)$ is the radial (orbital) wave function and $\mathcal{X}_{\lambda_{q} \lambda_{\bar{q}}}^{J J_{z}}$ is the spin-orbit wave function  obtained by the interaction-independent Melosh transformation. $n S (n P)$ refers to the various radially (orbitally) excited states and in this work, we have considered $n=1,2,3$ for the $S$ state and $n=1$ for the $P$ state. Covariant forms of  $\mathcal{X}$ for pseudo-scalar and vector mesons are respectively presented as \cite{Arifi:2022pal}

\begin{eqnarray}
\mathcal{X}_{\lambda_{q} \lambda_{\bar{q}}}^{00} &=&-\frac{1}{\sqrt{2} \mathcal{M}} \bar{u}_{\lambda_{q}}\left(p_{q}\right) \gamma_{5} v_{\lambda_{\bar{q}}}\left(p_{\bar{q}}\right),\\
\mathcal{X}_{\lambda_{q} \lambda_{\bar{q}}}^{1 J_{z}}&=&-\frac{1}{\sqrt{2} \mathcal{M}} \bar{u}_{\lambda_{q}}\left(p_{q}\right)\left[\notin\left(J_{z}\right)-\frac{\epsilon \cdot\left(p_{q}-p_{\bar{q}}\right)}{M_{0}+m_{q}+m_{\bar{q}}}\right] v_{\lambda_{\bar{q}}}\left(p_{\bar{q}}\right),
\end{eqnarray}
 where $\mathcal{M} \equiv \sqrt{M_{0}^{2}-\left(m_{q}-m_{\bar{q}}\right)^{2}}$. The polarisation vectors $\epsilon^{\mu}\left(J_{z}\right)=\left(\epsilon^{+}, \epsilon^{-}, \boldsymbol{\epsilon}_{\perp}\right)$ of the vector meson are given by 

\begin{align}
\epsilon^{\mu}( \pm 1) & =\left(0, \frac{2}{P^{+}} \boldsymbol{\epsilon}_{\perp}( \pm) \cdot \mathbf{P}_{\perp}, \boldsymbol{\epsilon}_{\perp}( \pm)\right), \\
\epsilon^{\mu}(0) & =\frac{1}{M_{0}}\left(P^{+}, \frac{-M_{0}^{2}+\mathbf{P}_{\perp}^{2}}{P^{+}}, \mathbf{P}_{\perp}\right),
\end{align}
where

\begin{equation}
\boldsymbol{\epsilon}_{\perp}( \pm 1)=\mp \frac{1}{\sqrt{2}}(1, \pm i).
\end{equation}

We define the radial wave function $\phi_{nS(nP)} (x, {\bf k_{\perp}})$ in Eq. (\ref{Twave}) for the $1S$, $2S$, $3S$ and $1P$ states as \cite{Dhiman:2019ddr,Hwang:2008qi}

\begin{eqnarray} \label{wave}
 \phi_{1 S}\left(x, \mathbf{k}_{\perp}\right)&=&\frac{1}{(\sqrt{\pi} \beta)^{3 / 2}} \exp \left(-\mathbf{k}^{2} / 2 \beta^{2}\right),\\ 
\phi_{2 S}\left(x, \mathbf{k}_{\perp}\right)&=&\frac{1}{(\sqrt{\pi} \beta)^{3 / 2}}\left(\frac{2 \mathbf{k}^{2}-3 \beta^{2}}{\sqrt{6} \beta^{2}}\right) \exp \left(-\mathbf{k}^{2} / 2 \beta^{2}\right), \\
\phi_{3 S}\left(x, \mathbf{k}_{\perp}\right)&=&\frac{1}{(\sqrt{\pi} \beta)^{3 / 2}}\left(\frac{15 \beta^{4}-20 \beta^{2} \mathbf{k}^{2}+4 \mathbf{k}^{4}}{2 \sqrt{30} \beta^{4}}\right) \exp \left(-\mathbf{k}^{2} / 2 \beta^{2}\right), \\
\phi_{1P}\left(x, \bf k_{\perp}\right) &=& \frac{1}{(\sqrt{\pi} \beta)^{3 / 2}} \left(\frac{\sqrt{2} k_{m}}{\beta} \right) \exp \left(-\mathbf{k}^{2} / 2 \beta^{2}\right), 
\end{eqnarray}
where, $(k_{m = \pm 1} = \mp (k_{\perp 1} \pm \iota k_{\perp 2})/ \sqrt{2}$, and  $k_{m = 0} = k_{z})$ and $m$ denotes the magnetic quantum number  obtained from the spherical harmonics. For this paper, we consider only $m = 0$ form of the wave function. $\beta$ is a variational parameter which we will be using in our model later in this paper. The radial wave functions of $\phi_{1S}$, $\phi_{2S}$, $\phi_{3S}$ and $\phi_{1P}$ can be normalized as

\begin{equation}
\int_{0}^{1} d x \int d^{2} \mathbf{k}_{\perp} \frac{\partial k_{z}}{\partial x}\left|\phi_{nS(nP)}\left(x, \mathbf{k}_{\perp}\right)\right|^{2}=1.
\end{equation}
We define the mass of the meson invariant under the influence of boost as 

\begin{equation}
    M^{2}_{0} = {\frac{\mathbf{k}_{\perp}^{2} + m_{q}^2}{x}} + {\frac{\mathbf{k}_{\perp}^{2} + m_{\overline{q}}^2}{1-x}}.
\end{equation}
The three-momentum ${\bf k} = (k_{z}, {\bf k_{\perp}})$ can be expressed as ${\bf k} = (x, \bf k_{\perp})$ \cite{Arifi:2022pal} using the following relation 
\begin{eqnarray}
    k_{z} = ( x - \frac{1}{2}) M_{0} + \frac{m_{\bar q}^{2} - m_{q}^{2}}{2 M_{0}}.
\end{eqnarray}
Consequently, the transformation of variables $[k_{z}, {\bf k_{\perp}}]$ $\rightarrow$ $[x, {\bf k_{\perp}}]$ is connected to the Jacobian variable in the following manner

\begin{equation}
\frac{\partial k_{z}}{\partial x}=\frac{M_{0}}{4 x(1-x)}\left[1-\frac{\left(m_{q}^{2}-m_{\tilde{q}}^{2}\right)^{2}}{M_{0}^{4}}\right]. 
\end{equation}

\section{Properties of the meson} \label{prop}

\subsection{Model Parameters}

In the LFQM,  the parameters required for our calculations are the quark masses $(m_{c}, m_{b})$ along with  two variational harmonic parameters $(\beta_{c \bar c}, \beta_{b \bar b})$.
We adopt the model parameters from the work of  Arifi \textit{et al.}  \cite{Arifi:2022pal} where they have obtained the input parameters using the variational principle. They have evaluated $ \langle \Psi | [H_{0} + V_{0}] \Psi \rangle$ while using a trial wave function depending on the values of $m$ and $\beta$ by further varying their values till they reach a minimum of the expectation value. These model parameters have been presented in Table \ref{tab:i}. These parameters yield good results for numerous attributes of the meson, for example, mass spectra, EMFFs, DAs and other related metrics. 

\begin{figure}[htbp]
	\centering
	\begin{subfigure}[b]{0.49\textwidth}
		\includegraphics[width=\textwidth]{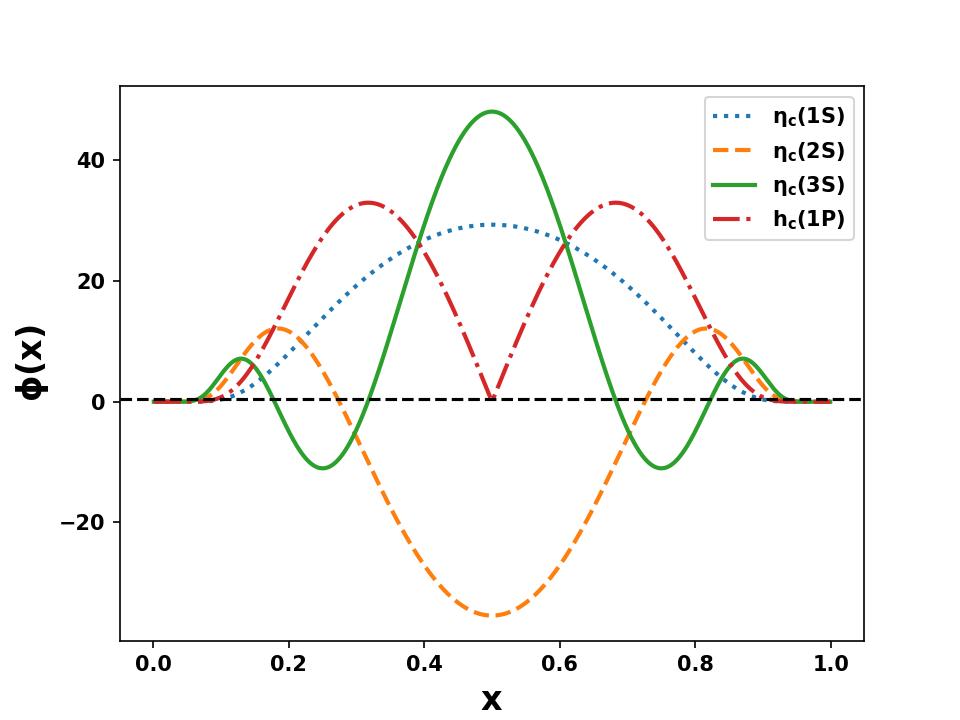}
		\caption{} \label{(1(a))}
	\end{subfigure}
	\begin{subfigure}[b]{0.49\textwidth}
		\includegraphics[width=\textwidth]{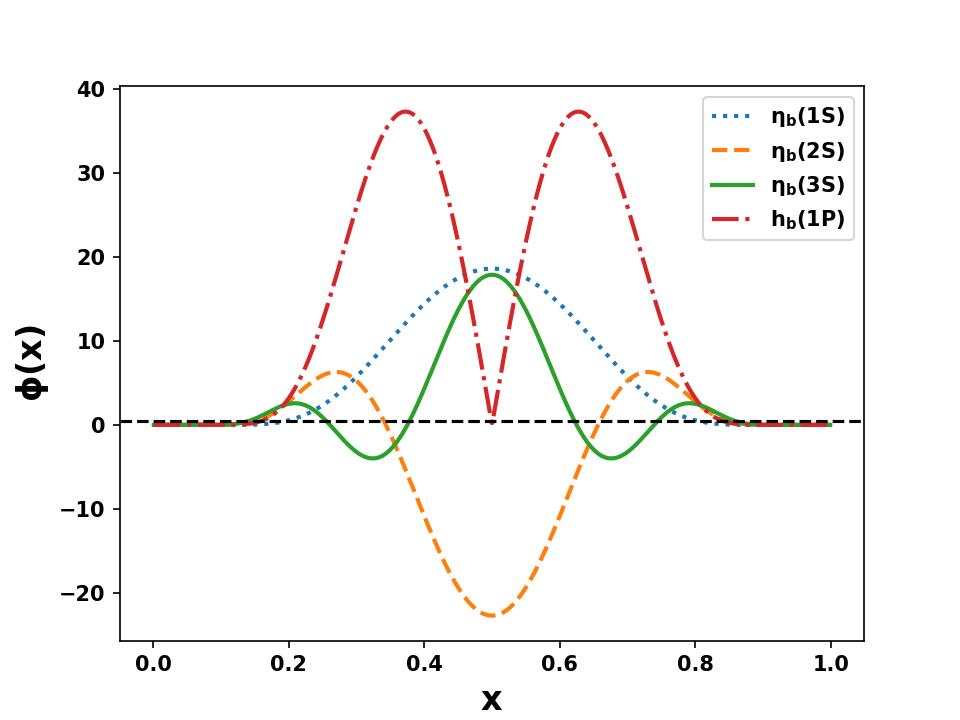}
		\caption{} \label{(1b)}
	\end{subfigure}
	\caption{2-D visualization of the excited state wave functions for spin-0 charmonium ($c \bar c$) (left side) and  bottomonium ($b  \bar b$) (right side) systems. \label{fig:1}}
	
\end{figure}

\begin{figure}[htbp]
	\centering
	\begin{subfigure}[b]{0.49\textwidth}
		\includegraphics[width=\textwidth]{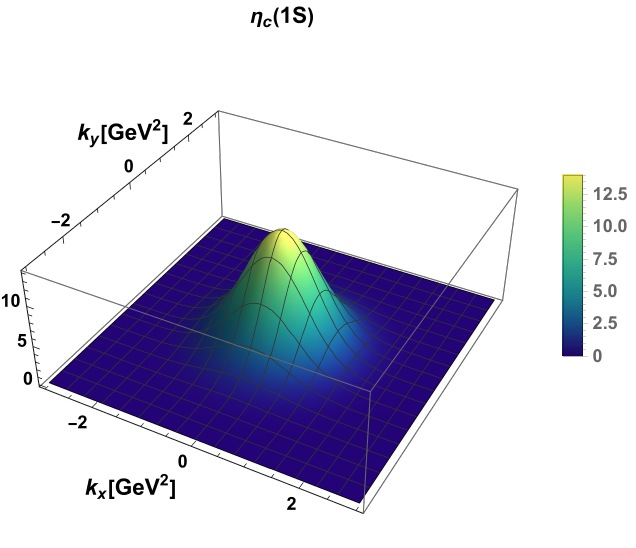}
		\caption{} \label{(a1111)}
	\end{subfigure}
	\begin{subfigure}[b]{0.49\textwidth}
		\includegraphics[width=\textwidth]{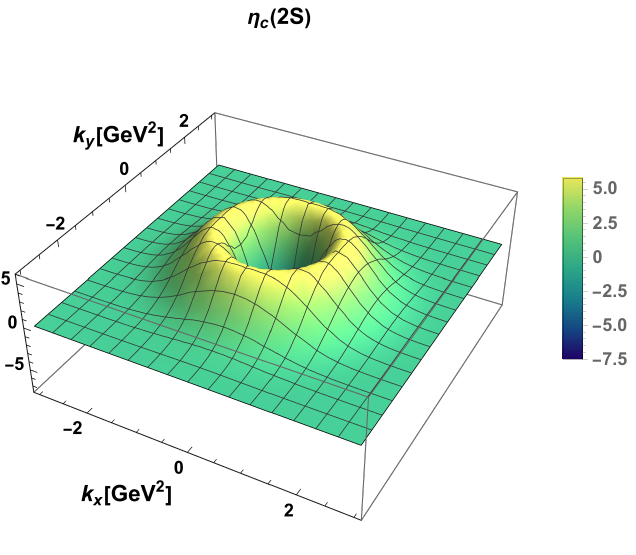}
		\caption{} \label{(b1111)}
	\end{subfigure}
	\qquad
	\begin{subfigure}[b]{0.5\textwidth}
		\includegraphics[width=\textwidth]{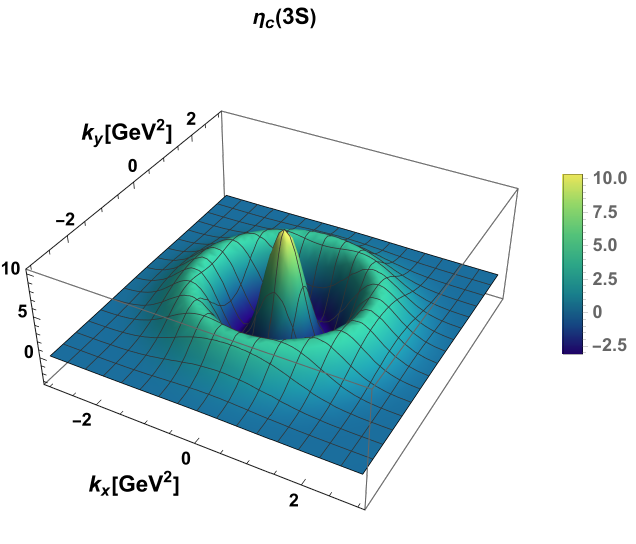}
		\caption{} \label{(c1111)}
	\end{subfigure}
	\begin{subfigure}[b]{0.49\textwidth}
	\end{subfigure}
	\caption{3-D visualization of the nodal structure contained within the radially excited states $1S$,  $2S$ and  $3S$ of the $\eta_{c}$ meson ((a), (b) and (c) respectively). \label{fig:4676}}
	
\end{figure}

In Fig. \ref{fig:1}, we have shown the momentum space wave functions for the states (defined in Sec. \ref{meth}) as a function of the longitudinal momentum fraction $x$ at a fixed transverse quark momenta $\bfk^{2} = 0.1$ GeV$^{2}$ for the charmonium and bottomonium mesons. We have considered only the quark distributions here since the quark and anti-quark masses are similar. These mesons tend to have a symmetric distribution in terms of the quark distributions with their anti-quark distributions being identical.
For a more intuitive understanding of the wave functions, we have provided 3-D representations of the wave functions as functions of $k_{x}$ and $k_{y}$, where $\bfk = \sqrt{k^{2}_{x} + k^{2}_{y}}$. We have presented these plots in Fig. \ref{fig:4676}. We have excluded the plot for the $1P$ state since its behavior is not very reflective of its wave nature. These plots portray a very beautiful understanding of the nodal structure of the wave functions. We notice that the $1S$ state has no nodes and has a Gaussian distribution which is as expected from its wave function construction. Similarly, for the $2S$ state we have a singular node whereas for the $3S$ state we have two subsequent nodes which are well represented in the figure. As we go to higher excited states, each state follows the trend of $(n-1)$ nodes where n is the principle quantum number of the given state.

\begin{table}[htbp] 
	\centering
	\begin{tabular}{|c|c|c|c|}
		\hline
	 $m_{c} (GeV)$ & $m_{b} (GeV)$ & $\beta_{c\overline{c}}$ & $\beta_{b\overline{b}}$\\
		\hline
		1.68&5.10&0.699&1.376\\
		\hline
	\end{tabular}
	\caption{Model parameters for LFQM \cite{Arifi:2022pal}.\label{tab:i}}
\end{table}

\subsection{Decay Constants}

The decay constants for a pseudo-scalar meson $\mathcal{P}$ can be defined from the axial-vector current's matrix components whereas for a vector meson $\mathcal{V}$  it can be defined from the vector current element. They are respectively expressed  as \cite{Arifi:2022pal,Choi:2021mni}
\be
\langle 0 | \overline{q} \gamma^{\mu} \gamma_{5} q | \mathcal{P} \rangle &=& \iota f_{\mathcal{P}} P^{\mu},\\
\langle 0 | \overline{q} \gamma^{\mu} q | \mathcal{V (\mathcal{P}, \lambda)} \rangle &=&  f_{\mathcal{V}} M \epsilon^{\mu} (\lambda).
\ee 

The explicit form of these current components for the case of pseudo-scalar mesons are denoted as $f^{(\pm)}_{\mathcal{P}}$ and expressed as

\begin{align}
	f_{\mathcal{P}}^{(\pm)} = \sqrt{6} \int dx \int \frac{d^{2} \mathbf{k}_{\perp}}{(2\pi)^{3}} \frac{\phi _{n S (n P)}(x, \mathbf{k}_{\perp})}{\sqrt{\mathcal{A}^{2} + \mathbf{k}_{\perp}^{2}}} \mathcal{O}^{(\pm)}_{\mathcal{P}}, 
\end{align}
where $\mathcal{A} = (1 - x) m_{q} + x m_{\overline{q}}$ and \begin{align}
	O_{\mathcal{P}}^{+} & =\mathcal{A}, \\
	O_{\mathcal{P}}^{-} & =\frac{\mathbf{k}_{\perp}^{2} \mathcal{A}^{\prime}+m_{q} m_{\bar{q}} \mathcal{A}}{x(1-x) M_{0}^{2}},
\end{align}
with $\mathcal{A}^{\prime}=\mathcal{A}\left(m_{q} \leftrightarrow m_{\bar{q}}\right)$. For the case of pseudo-scalar mesons, the $``+"$ and $``-"$ current components correspond to the generators for the decay constants of ground state ($1S$) and first excited state ($2S$) state mesons  \cite{Arifi:2022pal, Choi:2021mni} (previously proven to be identical). We extend this study to the cases of pure $3S$ and $1P$ states. 

Similarly, for the vector meson case, the  $``+"$ current component with longitudinal polarization $\epsilon$ (0) is equivalent to the $\perp$ current component carrying the transverse polarization $\epsilon$ ($\pm$) and they have also proven to be completely identical \cite{Choi:2013mda}. We denote the vector decay constant $f_{\mathcal{V}}$ for the $\perp$ and $``+"$ current components as $f^{\perp}_{\mathcal{V}}$ and  $f^{+}_{\mathcal{V}}$, respectively. We express them quantitatively as 


\begin{equation}
f_{\mathcal{V}}^{(+,\perp)} = \sqrt{6} \int dx \int \frac{d^{2} \mathbf{k}_{\perp}}{(2\pi)^{3}} \frac{\phi _{n S (n P)}(x, \mathbf{k}_{\perp})}{\sqrt{\mathcal{A}^{2} + \mathbf{k}_{\perp}^{2}}} \mathcal{O}^{(+ \perp)}_{\mathcal{V}}, 
\end{equation}
where
\begin{align}
O_{\mathcal{V}}^{+} & =\mathcal{A} + \frac{2 \bfk^{2}}{G_{LF}}, \\
O_{\mathcal{V}}^{\perp} & = \frac{1}{M_{0}} \left [ \frac{\bfk^{2} + \mathcal{A}^{2}}{2x(1-x)} - \bfk^{2} + \frac{(m_{q} + m_{\bar q})}{G_{LF}} \bfk^{2} \right],
\end{align}
and $G_{LF} = M_{0} + m_{q} + m_{\bar q}$.  The decay constants $f^{+}_{\mathcal{V}}$ and $f^{\perp}_{\mathcal{V}}$ computed using the basis light-front quantization (BLFQ) technique \cite{Li:2018uif} differed based on the current components considered in their work which was attributed to the model's violation of its rotational symmetry. However, in LFQM, we have been able to successfully calculate that $f_{\mathcal{P}}$ = $f^{+}_{\mathcal{P}}$ = $f^{-}_{\mathcal{P}}$ and $f_{\mathcal{V}}$ = $f^{+}_{\mathcal{V}}$ = $f^{\perp}_{\mathcal{V}}$ for all the radially excited states of the mesons.

\begin{table}[h]
\footnotesize
	\centering
	\begin{tabular}{|c|c|c|c|c|}
		\hline    
		\hline & $f_{\eta_{c}(1 S)}$ & $f_{J / \Psi(1 S)}$ & $f_{\eta_{b}(1 S)}$ & $f_{\Upsilon(1 S)}$ \\
		\hline Our results & 356 & 403 & 648 & 688 \\
		Expt. \cite{ParticleDataGroup:2020ssz} & $335 \pm 75$ & $407 \pm 5$ & — & $689 \pm 5$ \\
		Lattice \cite{McNeile:2012qf,Davies:2010ip,Donald:2012ga,Colquhoun:2014ica,Becirevic:1998ua} & $394.7 \pm 2.4$ & $405 \pm 6$ & $667_{-2}^{+6}$ & $649 \pm 31$ \\
		RQM 
		Sum rules \cite{Becirevic:2013bsa} & $387 \pm 7$ & $418 \pm 9$ & - & - \\
		BS \cite{Cvetic:2004qg}
		& $292 \pm 25$ & $459 \pm 28$ & - & $496 \pm 20$ \\
		BS \cite{Binosi:2018rht} & 385 & - & 709 & - \\
		LFQM (CJ) \cite{Choi:2009ai} & 326 & 360 & 507 & 529 \\
		\multirow[t]{2}{*}{ LFQM (CJ2) \cite{Choi:2015ywa} } & 353 & 361 & 605 & 611 \\
		\hline
		\hline & $f_{\eta_{c}(2 S)}$ & $f_{\Psi^{\prime}(2 S)}$ & $f_{\eta_{b}(2 S)}$ & $f_{\Upsilon(2 S)}$ \\
		\hline Our calculation & 318 & 420 & 671 & 771 \\
		Expt. \cite{ParticleDataGroup:2020ssz} & - & $294(5)$ & - & $497(5)$ \\
		Lattice \cite{Colquhoun:2014ica} & - & - & - & $481(39)$ \\
		BLFQ \cite{Li:2017mlw}
		${ }^{a}$ & 298 & 312 & 525 & 520 \\
		\multirow[t]{2}{*}{LFD \cite{Li:2021cwv}} & - & $288(6)$ & - & - \\
		\hline
		\hline & $f_{\eta_{c}(3 S)}$ & $f_{\Psi^{\prime}(3 S)}$ & $f_{\eta_{b}(3 S)}$ & $f_{\Upsilon(3 S)}$ \\
		\hline Our calculation & 239 & 357 & 586 & 724 \\
		RHCM \cite{Hoque:2020qak} & 420 & - & 711 & - \\
		\multirow[t]{2}{*}{BS \cite{Zhou:2020bnm}} &- & 375 &- &- \\
		
		\hline
		\hline & $f_{h_{c}(1P)}$ & $f_{\chi_{c0}(1 P)}$ & $f_{h_{b}(1P)}$ & $f_{\chi_{b0}(1 P)}$ \\
		\hline Our calculation & 373 & 421 & 701 & 745 \\
		Sum rules \cite{Wang:2012gj} & $490$ & - & 657 & - \\
		BS \cite{Zhou:2020bnm} &- &239 &- &- \\
		\hline \hline
	\end{tabular}
	\caption{Decay constants of heavy quarkonia for  various radially and orbitally excited states. We compare our results with the available theoretical and experimental data.}
	\label{tab:my_label}
\end{table}

In Table \ref{tab:my_label}, we have presented  our theoretical calculations for decay constants of heavy pseudo-scalar mesons, $\eta_{c}$ and $\eta_{b}$, along with heavy vector mesons $J /\Psi$ and $\Upsilon$ for their  ground state and excited states. For the sake of comparison, we have provided the available experimental data along with some other theoretical calculations like the lattice simulations \cite{McNeile:2012qf,Davies:2010ip,Donald:2012ga,Colquhoun:2014ica,Becirevic:1998ua} and QCD sum rules \cite{Gelhausen:2014jea}. Since we have adopted the model parameters from Ref. \cite{Arifi:2022pal}, our results for the $1S$ and $2S$ states are identical. From our calculations it is observed that the decay constants have an increasing trend as we consider the higher radially excited states for a particular meson. For the case of $\phi_{nS}$, the decay constants follow the trend $f_{1 S} < f_{2 S} < f_{3 S}$ which is in violation of the empirical hierarchy discussed in Ref. \cite{Arifi:2022pal} stating that $f_{1 S} > f_{2 S}$. This is due to the fact that they have mixed the 1S and 2S states in order to satisfy the hierarchy. This further makes us believe that mixing even higher excited states would help us obtain better results and uphold the empirical hierarchy.

\subsection{Distribution Amplitudes (DAs)}

In LFQM, one can extract the DA $\phi_{\mathcal{P (V)}}(x, \mu)$ of a meson from its LF wave function by integrating over the transverse momenta $\bfk$ of the quarks in the meson. DAs provide the probability amplitudes for locating the hadron in a state with a minimal transverse momentum separation and a minimum amount of Fock elements, identified by an ultraviolet (UV) cutoff of $\mu \gtrsim 1$ GeV. The QCD evolution equation \cite{Lepage:1980fj} provides the $\mu$-scale dependence which can be determined perturbatively. However, the DAs can be extracted using contributions from the non-perturbative regimes of the LFQM  at certain low scale values. Moreover, LFQM includes a damping Gaussian component which allows the computation of indefinite integrals with no loss in accuracy. Pseudo-scalar and vector meson quark DAs are correlated with decay constants derived from the $``+"$ current constituents by the relation \cite{Choi:2007yu}

\begin{equation}
\int_{0}^{1} \phi_{\mathcal{P (V)}} (x, \mu) d x=\frac{f_{\mathcal{P (V)}}^{(+)}}{2 \sqrt{6}}.
\end{equation} 
We define the normalized quark DA as ${\phi''}_{\mathcal{P(V)}} (x, \mu)=$ $\left(2 \sqrt{6} / f_{\mathcal{P(V)}}^{(+)}\right) \phi_{\mathcal{P(V)}} (x, \mu)$ so that

\begin{equation}
\int_{0}^{1} {\phi''}_{\mathcal{P (V)}}(x, \mu) d x = 1.
\end{equation}

It is typically possible to expand the quark DAs using Gegenbaur polynomials $C_{n}^{3/2}$ as ${\phi''}(x, \mu)={\phi''}_{\text {asym}}(x)\left[1+\sum_{n=1}^{\infty} e_{n}(\mu) C_{n}^{3 / 2}(\xi)\right]$, where the asymptotic DA is defined as ${\phi''}_{\text {asym}}(x) = 6 x (1-x)$ and $e_{n}(\mu)$ are the Gegenbaur moments \cite{Lepage:1980fj,RuizArriola:2002bp, Mueller:1994cn}. With $n > 0$, the Gegenbauer moments describe the DAs' divergence from the asymptotic form. As an alternative, the longitudinal momentum's expectation value intricately connected to $e_{n}(\mu)$ can be defined as \cite{Choi:2007yu}
\begin{equation}\label{xi}
\left\langle\xi^{n}\right\rangle=\int_{0}^{1} dx \vspace{2mm} \xi^{n} \vspace{2mm} {\phi''}_{\mathcal{P (V)}}(x, \mu),
\end{equation}
where  $\xi=x-(1-x)=2 x-1$.


\begin{figure}[htbp]
	\centering
	\begin{subfigure}[b]{0.49\textwidth}
		\includegraphics[width=\textwidth]{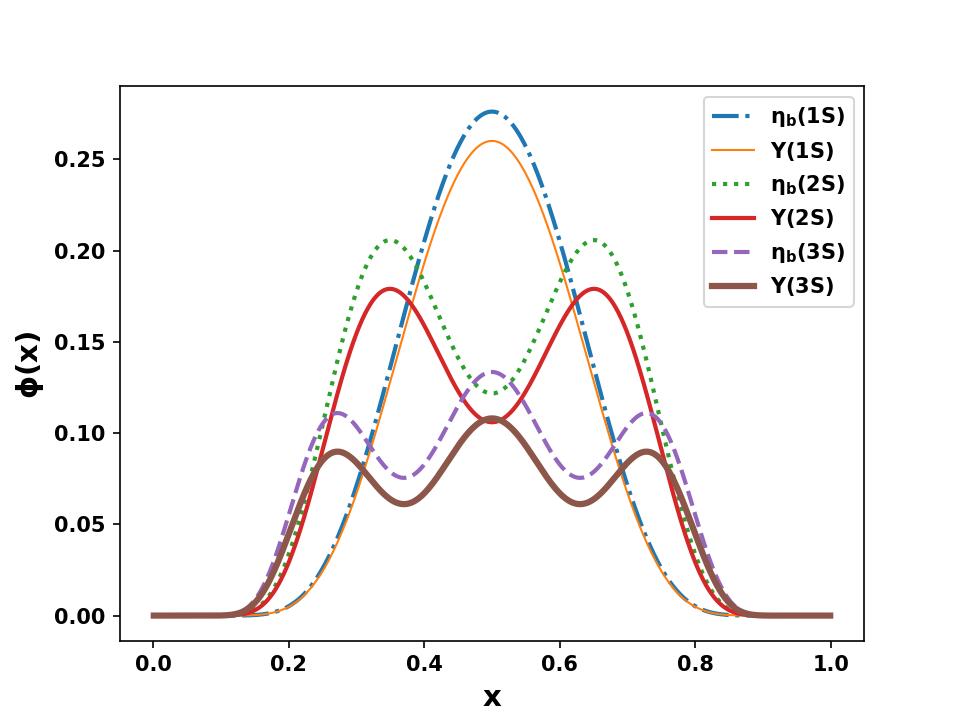}
		\caption{} \label{(a11)}
	\end{subfigure}
	\begin{subfigure}[b]{0.49\textwidth}
		\includegraphics[width=\textwidth]{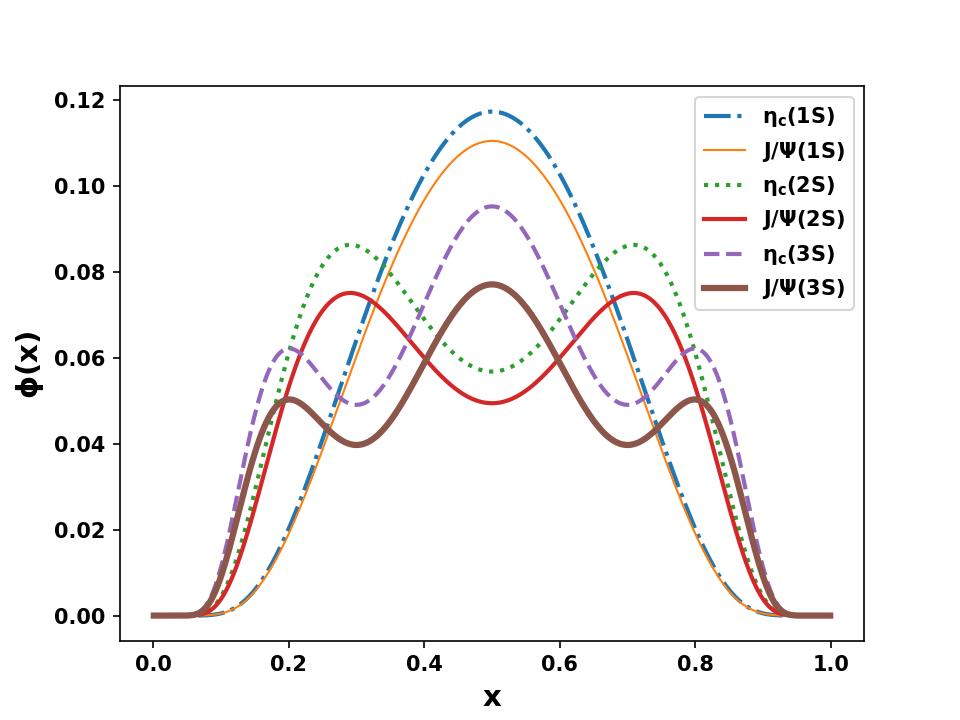}
		\caption{} \label{(b11)}
	\end{subfigure}
	\qquad
	\begin{subfigure}[b]{0.49\textwidth}
		\includegraphics[width=\textwidth]{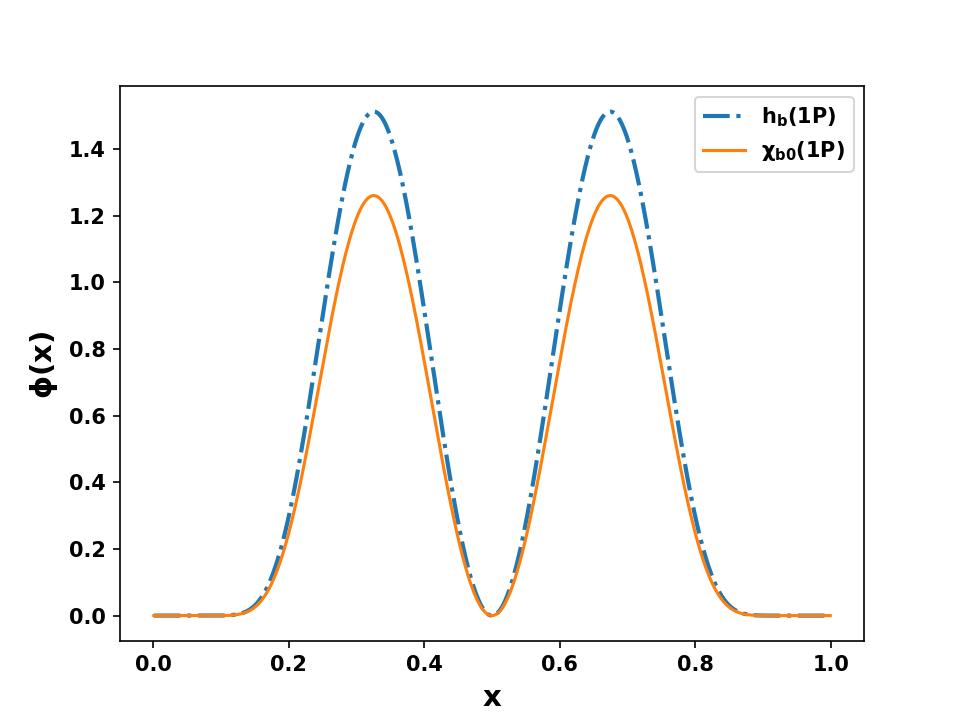}
		\caption{} \label{(c11)}
	\end{subfigure}
	\begin{subfigure}[b]{0.49\textwidth}
		\includegraphics[width=\textwidth]{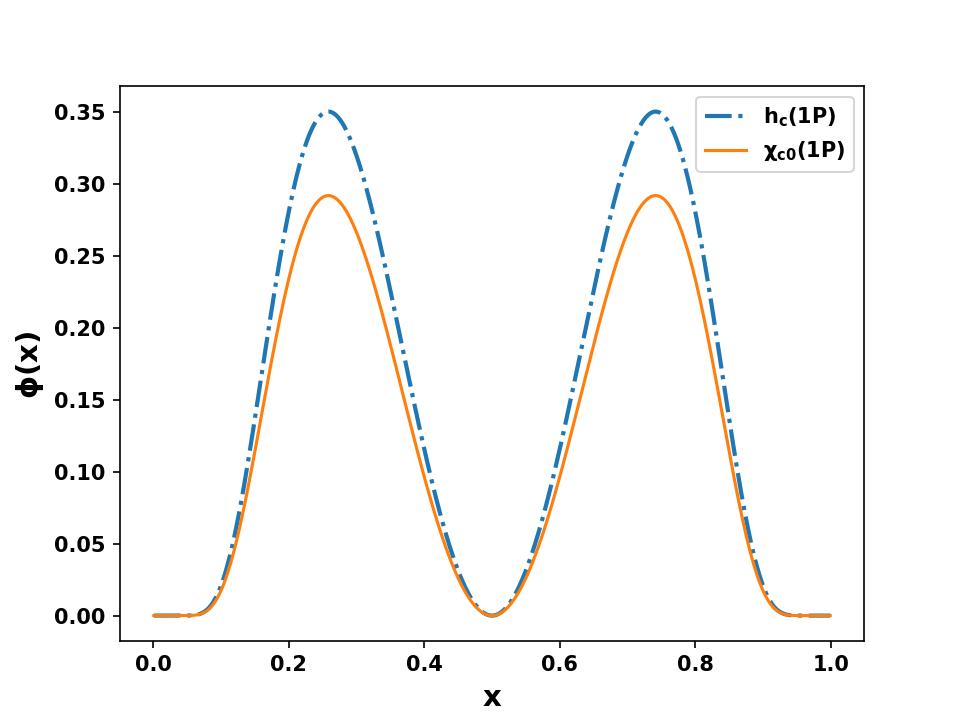}
		\caption{} \label{(d11)}
	\end{subfigure}
	\caption{We compare the DAs of the different excited states for the pseudo-scalar and vector bottomonia ($b \bar b$) and charmonia ($c \bar c$) systems. The radially excited states are compared in the upper panel and the orbitally excited state ($1P$) in the lower panel. \label{fig:456}}
	
\end{figure}

In Fig. \ref{fig:456}, we have compared the DAs of spin-0 and spin-1 mesons for the radially excited states and the orbitally excited states separately. In the left panel we show the comparisons for the states of bottomonia ($b \bar{b}$) system and on the right panel we compare the charmonia ($c \bar{c}$) systems. We choose the convention where the quark carries the longitudinal momentum fraction $x$ and the anti-quark carries the fraction $1-x$. Since we are considering the charmonia ($c \bar c$) and bottomonia ($b \bar b$) systems, our quark and anti-quark pair have identical masses. The plot indicates that the vector mesons in LFQM model and the pseudo-scalar mesons in the $1S$ state are quite comparable. However, when we consider the higher excited states, we have observed that the pseudo-scalar and vector meson DAs have a more pronounced difference in their amplitudes. This discrepancy seems to increase as we go to higher excited states. The DA structure for the bottomonia is narrower than the charmonia even though both the heavy quarkonia DAs are symmetric under $x \rightarrow 1-x$. We also observe that the amplitude for the bottomonium DA is larger than the charmonium one. This difference, in our view, results from the enormous mass of the $b$ quark and its significantly larger $\beta$ parameter as compared to the $c \bar c$ system.  For the $1P$ state, we have observed that the wave function has two amplitude peaks and they tend to have a minimum at $x = 0.5$. This is because our chosen wave function has a $x - \frac{1}{2}$ term in the numerator when we substitute the explicit form of the $k_{z}$ into the orbital wave function.
\begin{figure}[htbp]
	\centering
	\begin{subfigure}[b]{0.35\textwidth}
		\includegraphics[width=\textwidth]{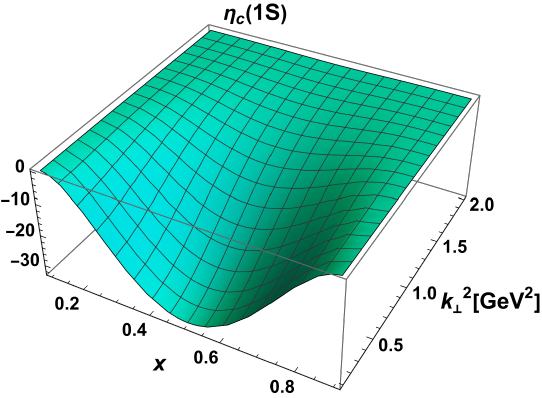}
		\caption{} \label{(a1)}
	\end{subfigure}
    \begin{subfigure}[b]{0.35\textwidth}
		\includegraphics[width=\textwidth]{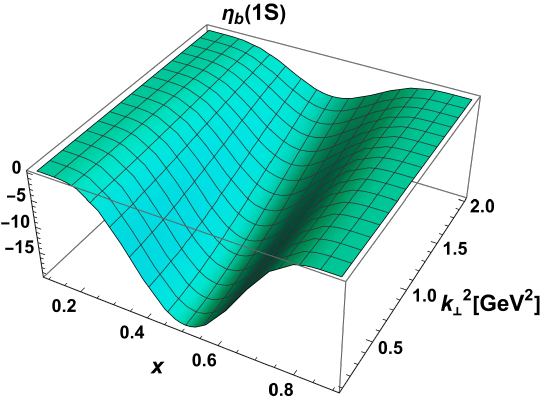}
		\caption{} \label{(a111)}
    \end{subfigure}
	\begin{subfigure}[b]{0.35\textwidth}
		\includegraphics[width=\textwidth]{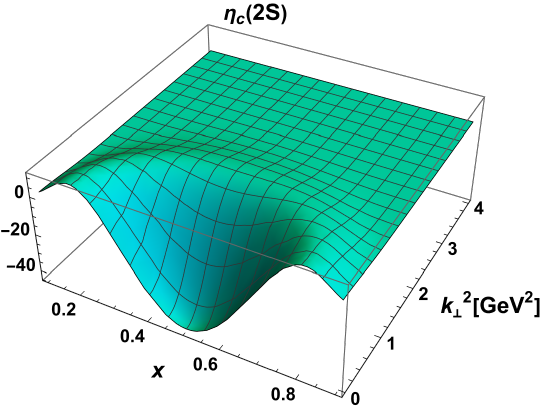}
		\caption{} \label{(b1)}
	\end{subfigure}
    \begin{subfigure}[b]{0.35\textwidth}
		\includegraphics[width=\textwidth]{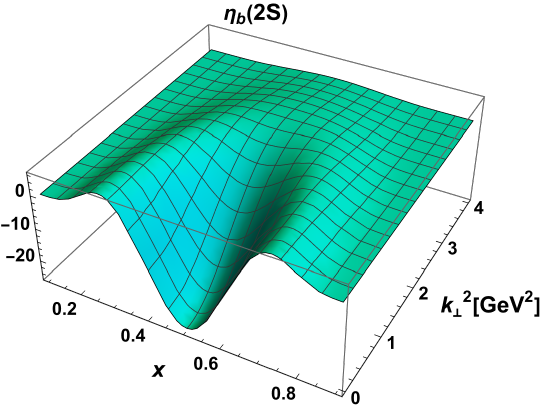}
		\caption{} \label{(b111)}
	\end{subfigure}
	\qquad
	\begin{subfigure}[b]{0.35\textwidth}
		\includegraphics[width=\textwidth]{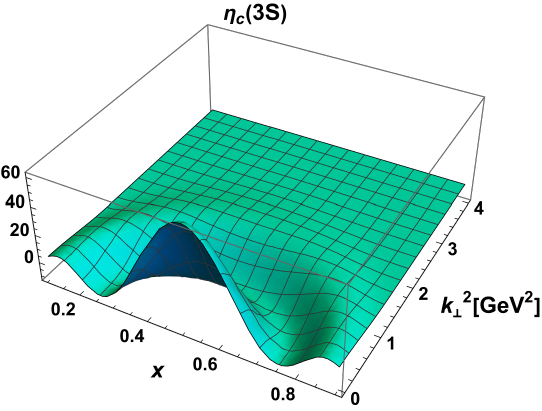}
		\caption{} \label{(c1)}
	\end{subfigure}
    \begin{subfigure}[b]{0.35\textwidth}
		\includegraphics[width=\textwidth]{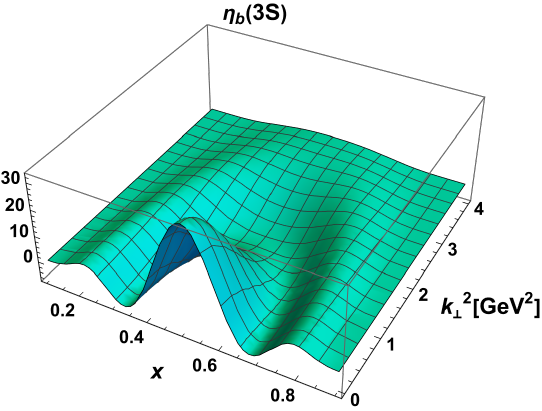}
		\caption{} \label{(c111)}
	\end{subfigure}
	\begin{subfigure}[b]{0.35\textwidth}
		\includegraphics[width=\textwidth]{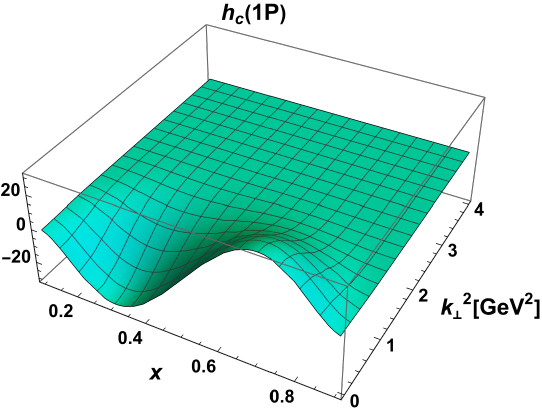}
		\caption{} \label{(d1)}
	\end{subfigure}
    \begin{subfigure}[b]{0.35\textwidth}
		\includegraphics[width=\textwidth]{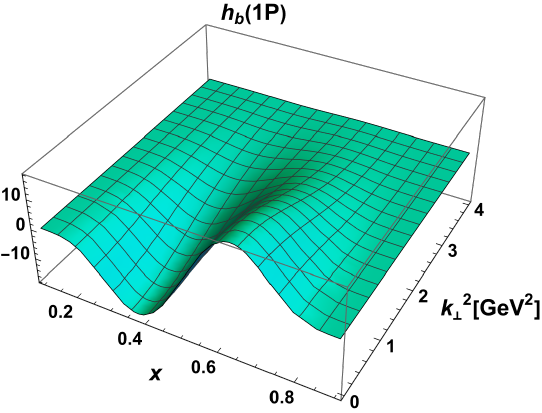}
		\caption{} \label{(d111)}
	\end{subfigure}
	\caption{3D plots of $\Psi_{\eta_{c}} (x,\bfk)$ and $\Psi_{\eta_{b}} (x,\bfk)$ for the radially and orbitally excited states of $\eta_{c}$ meson (left panel) and of $\eta_{b}$ meson (right panel). \label{fig:5}}
\end{figure}

The DAs for the heavy quarkonia can be normalized and rewritten as
\begin{equation} \label{norm}
	{\phi''}_{\mathcal{P(V)}} (x, \mu) = \int_{0}^{|\bfk| < \mu} d^{2} \bfk \psi''_{\mathcal{P(V)}} (x, \bfk),
\end{equation}
where $\psi''_{\mathcal{P(V)}} (x, \bfk)$ is the LF wave function associated with ${\phi''}_{\mathcal{P(V)}} (x, \mu)$. In Fig. \ref{fig:5}, we have presented the 3-D plots for $\psi''_{\eta_{c}} (x, \bfk)$ and $\psi''_{\eta_{b}} (x, \bfk)$ of their respected excited states. It is inferred from Eq. \ref{norm} that the normalized DAs ${\phi''}_{\mathcal{P(V)}} (x, \mu)$ shown in Fig. \ref{fig:456} are calculated by integrating $\psi''_{\mathcal{P(V)}} (x, \bfk)$ over the transverse momentum $\bfk$. We find that the UV cutoffs or energy scale $\mu$ correlate to $|\bfk| \rightarrow \infty$ in LFQM using the Gaussian wave functions. They are approximately equal to 2 GeV for $\eta_{c}$ and 4 GeV for $\eta_{b}$. This clearly indicates that the wave functions for the heavy-heavy systems have longer transverse momentum tails as compared to the  light-heavy systems. This has also been concluded in Ref. \cite{Arifi:2022pal}. On the other hand, the wave functions for systems having a heavy quark and a light anti-quark pair exhibit larger negative troughs than those for systems having both heavy quark and anti-quark pair. This characteristic of the LF wave function explains the presence of negative areas exclusively in the DAs for the higher excited state heavy-light system.

\begin{table}[]
    \centering
    \noindent
\begin{tabularx}{\textwidth}{|X|X|X|X|X|X|}
\hline \hline
($1S$) & $\eta_{c}$ & $J/\Psi$ & $\eta_{b}$ & $\Upsilon$ \\
\hline

$\langle \xi^{2} \rangle$ & 0.089 & 0.089 & 0.051 & 0.051 \\
$\langle \xi^{4} \rangle$ & 0.019 & 0.019 & 0.006 & 0.007 \\
$\langle \xi^{6} \rangle$ & 0.0056 & 0.0056 & 0.0013 & 0.0013 \\
\hline
\hline
($2S$) & $\eta_{c}$ & $\Psi$ & $\eta_{b}$ & $\Upsilon$ \\
\hline

$\langle \xi^{2} \rangle$ & 0.168 & 0.167 & 0.105 & 0.105\\
$\langle \xi^{4} \rangle$ & 0.049 & 0.050 & 0.020 & 0.020 \\
$\langle \xi^{6} \rangle$ & 0.018 & 0.018 & 0.005 & 0.005 \\
\hline
\hline
($3S$) & $\eta_{c}$ & $\Psi$ & $\eta_{b}$ & $\Upsilon$ \\
\hline

$\langle \xi^{2} \rangle$ & 0.169& 0.169 & 0.066 & 0.13 \\
$\langle \xi^{4} \rangle$ & 0.059& 0.059 & 0.016 & 0.031 \\
$\langle \xi^{6} \rangle$ & 0.025 & 0.025 & 0.005 & 0.009 \\
\hline
\hline
($1P$) & $h_{c}$ & $\chi_{c0}$ & $h_{b}$ & $\chi_{b0}$ \\
\hline

$\langle \xi^{2} \rangle$ & 0.166 & 0.089 & 0.091 & 0.051 \\
$\langle \xi^{4} \rangle$ & 0.045 & 0.019 & 0.016 & 0.006 \\
$\langle \xi^{6} \rangle$ & 0.015 & 0.005 & 0.004 & 0.0013 \\
\hline
\hline
\end{tabularx}
    \caption{$\xi$-moment upto $n=6$ for the excited states of pseudo-scalar and vector quarkonia.}
    \label{xi_mom}
\end{table}

\begin{table}[]
\scriptsize
    \centering
    \noindent
\begin{tabularx}{\textwidth}{|X|X|X|X|X|X|X|X|X|X|}
\hline
\hline
 &  & LFQM & NRQCD \cite{Bodwin:2006dn} & QCDSR \cite{Braguta:2006wr,Braguta:2007fh,Braguta:2007tq} & DSE \cite{Ding:2015rkn} & BLFQ \cite{Li:2017mlw} & Buchmuller-Tye model \cite{Buchmuller:1980su}   \\
\hline
 & $\langle \xi^{2} \rangle$ & 0.089 & 0.075 & 0.070 & 0.10 & 0.096 & 0.086\\
$\eta_{c} (1S)$ & $\langle \xi^{4} \rangle$ & 0.019 & 0.010 & 0.012 & 0.032 & 0.019 & 0.020\\
 & $\langle \xi^{6} \rangle$ & 0.0056 & 0.0017 &  0.0032 & 0.015 & 0.0036 & 0.0066 \\
\hline
 & $\langle \xi^{2} \rangle$ & 0.168 & 0.22 & 0.18 &  & 0.157 & 0.16\\
$\eta_{c} (2S)$ & $\langle \xi^{4} \rangle$ & 0.049 & 0.085 & 0.051 &  & 0.043 & 0.042\\
 & $\langle \xi^{6} \rangle$ & 0.018 & 0.039 &  0.017 &  & 0.013 & 0.015 \\
\hline
 & $\langle \xi^{2} \rangle$ & 0.089 & 0.075 & 0.070 & 0.039 & 0.096 & 0.086 \\
$J/\Psi (1S)$ & $\langle \xi^{4} \rangle$& 0.019 & 0.010 & 0.012 & 0.038 & 0.021 & 0.020 \\
 & $\langle \xi^{6} \rangle$ & 0.0056 & 0.0017 & 0.0031 &  7.3 $\times$ $10^{-4}$ & 0.0060 & 0.0066 \\
\hline
 & $\langle \xi^{2} \rangle$ & 0.051 & &  & 0.070 & 0.052 &  \\
$\eta_{b} (1S)$  & $\langle \xi^{4} \rangle$ & 0.006 & & & 0.015 & 0.081 &  \\
 & $\langle \xi^{6} \rangle$ & 0.0013 &  &  & 0.042 & 0.0020 & \\
\hline
 & $\langle \xi^{2} \rangle$ & 0.105 & &  &  & 0.082 &  \\
$\eta_{b} (2S)$  & $\langle \xi^{4} \rangle$ & 0.020 & & &  & 0.013 &  \\
 & $\langle \xi^{6} \rangle$ & 0.005 &  &  & & 0.003 & \\
\hline
 & $\langle \xi^{2} \rangle$ & 0.051 & &  &  0.014& 0.047 &  \\
$\Upsilon (1S)$  & $\langle \xi^{4} \rangle$ & 0.007 & & & 4.3 $\times 10^{-4}$  & 0.0066 &  \\
 & $\langle \xi^{6} \rangle$ & 0.0013 &  &  & 4.4 $\times 10^{-5}$ & 0.0014 & \\
\hline
\hline
\end{tabularx}
    \caption{Comparison of some of our LFQM $\xi$-moment results in the $1S$ and $2S$ states with different theoretical models and sum rules.}
    \label{xicomp}
\end{table}

In Table \ref{xi_mom}, we have presented the $\xi$-moments for various excited states of the heavy quarkonia using the relation defined in Eq. (\ref{xi}). For our charmonia ($c \bar c$) and bottomonia ($b \bar b$) case, the odd-moments vanish due to their DAs having a symmetric waveform. Therefore, we  have provided the results only for the even-moments upto $n = 6$. We note that our choice is due to the reasoning provided in Ref. \cite{Braguta:2006wr} where the authors have mentioned that there is a larger contribution from the end points ($x \sim 0$ and $x \sim 1$) as we consider higher powers of the moment. At these points in space, the mobility of the quark-antiquark pair becomes relativistic and we cannot consider potential models to properly explain their behavior. Hence, it is not possible to calculate the higher powers of the moment within our potential model. Therefore, we chose only the first few moments up to $n = 6$. For brevity, we have compared some of our results with other available theoretical models and have summarized them in Table \ref{xicomp}.

\subsection{Electromagnetic Form Factors (EMFFs)}

The charge radii and EMFFs of heavy pseudo-scalar meson have also been computed in our model. We have followed the Drell-Yan-West frame ($q^{+} = 0$) with $\mathbf{q}_{\perp}^{2}=Q^{2}=-q^{2}$ \cite{Arifi:2022pal}. Owing to the existence of several formulations for exact FFs in the light-front regime, we believe that this is a clear benefit in the LFQM. 
For the ``$+$"-component of the current $J^{\mu}$, the pseudo-scalar meson's charge FF can be represented as \cite{Choi:1997iq}
\begin{equation}
\mathcal{F}\left(Q^{2}\right)=e_{q} I^{+}\left(Q^{2}\right)+e_{\bar{q}} I^{+}\left(Q^{2}\right),
\end{equation}
where $e_{q}\left(e_{\bar{q}}\right)$ is the electric charge of quark (anti-quark), and
\begin{equation}
I^{+}\left(Q^{2}\right)=\int_{0}^{1} d x \int \frac{d^{2} \mathbf{k}_{\perp}}{2(2 \pi)^{3}} \phi\left(x, \mathbf{k}_{\perp}\right) \phi^{*}\left(x, \mathbf{k}_{\perp}^{\prime}\right) \\
\times
\frac{\mathcal{A}^{2}+\mathbf{k}_{\perp} \cdot \mathbf{k}_{\perp}^{\prime}}{\sqrt{\mathcal{A}^{2}+\mathbf{k}_{\perp}^{2}} \sqrt{\mathcal{A}^{2}+\mathbf{k}_{\perp}^{\prime 2}}},
\end{equation}
where $\mathbf{k}_{\perp}^{\prime}=\mathbf{k}_{\perp}+(1-x) \mathbf{q}_{\perp}$. The charge radius is thus given as
\begin{equation}
\left\langle r^{2}\right\rangle=-\left.6 \frac{d \mathcal{F}\left(Q^{2}\right)}{d Q^{2}}\right|_{Q^{2}=0}.
\end{equation}

\begin{figure}[htbp]
	\centering
	\begin{subfigure}[b]{0.49\textwidth}
		\includegraphics[width=\textwidth]{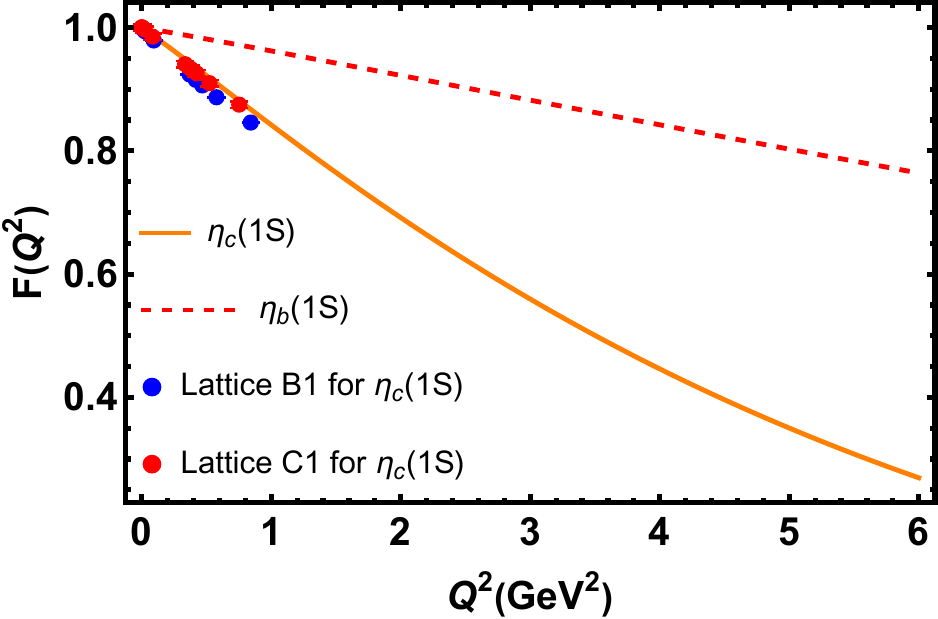}
		\caption{} \label{(a1)}
	\end{subfigure}
	\begin{subfigure}[b]{0.5\textwidth}
		\includegraphics[width=\textwidth]{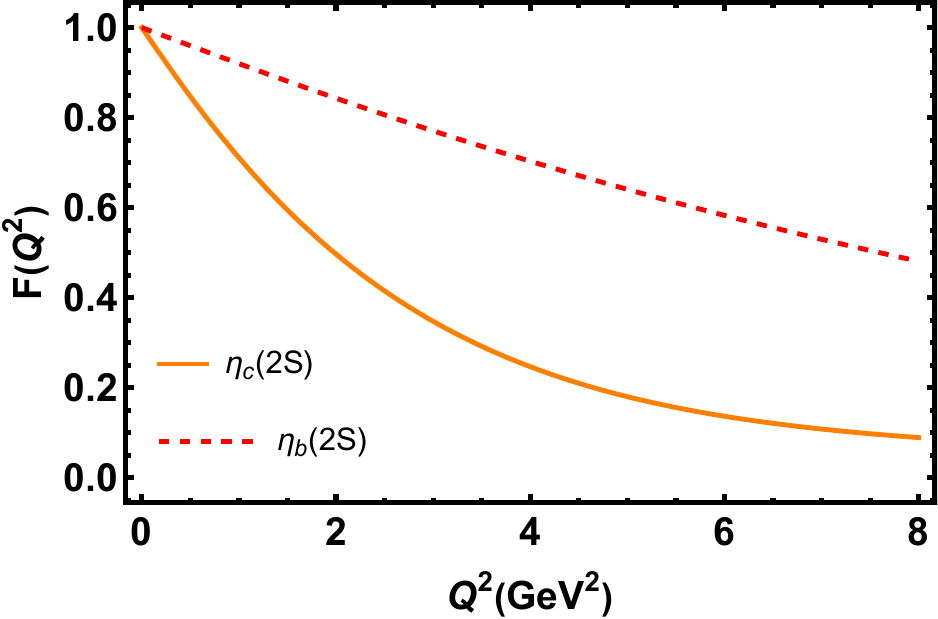}
		\caption{} \label{(b1)}
	\end{subfigure}
	\qquad
	\begin{subfigure}[b]{0.49\textwidth}
		\includegraphics[width=\textwidth]{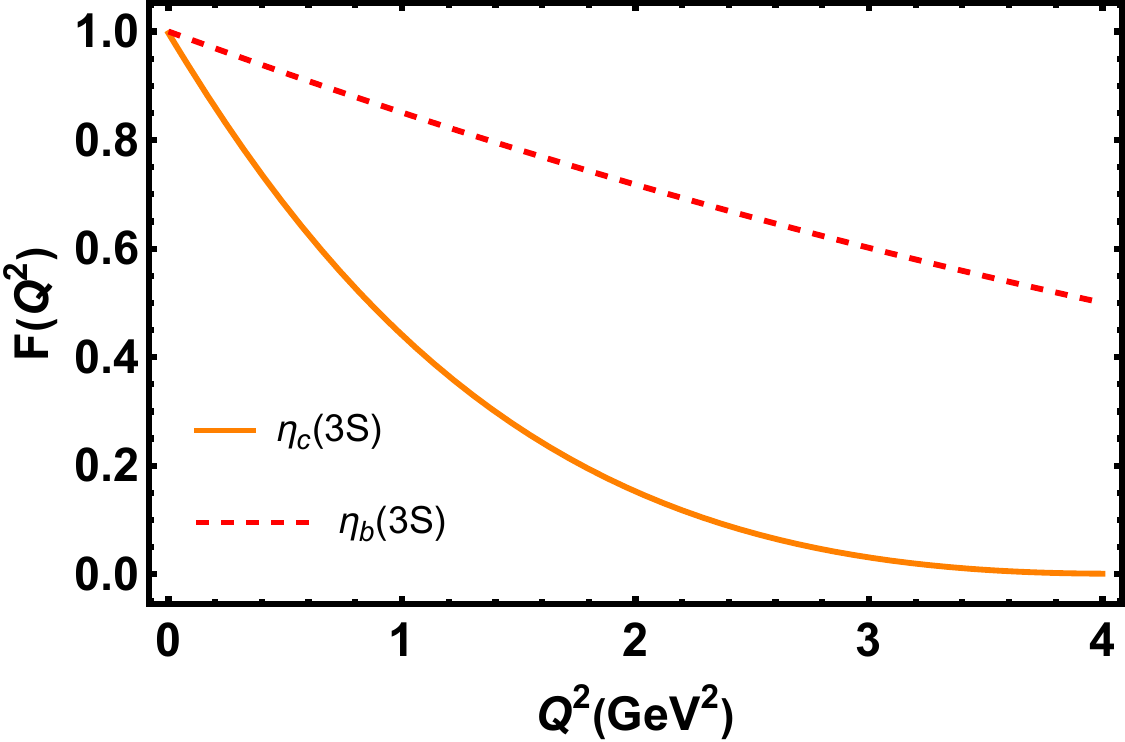}
		\caption{} \label{(c1)}
	\end{subfigure}
	\begin{subfigure}[b]{0.49\textwidth}
		\includegraphics[width=\textwidth]{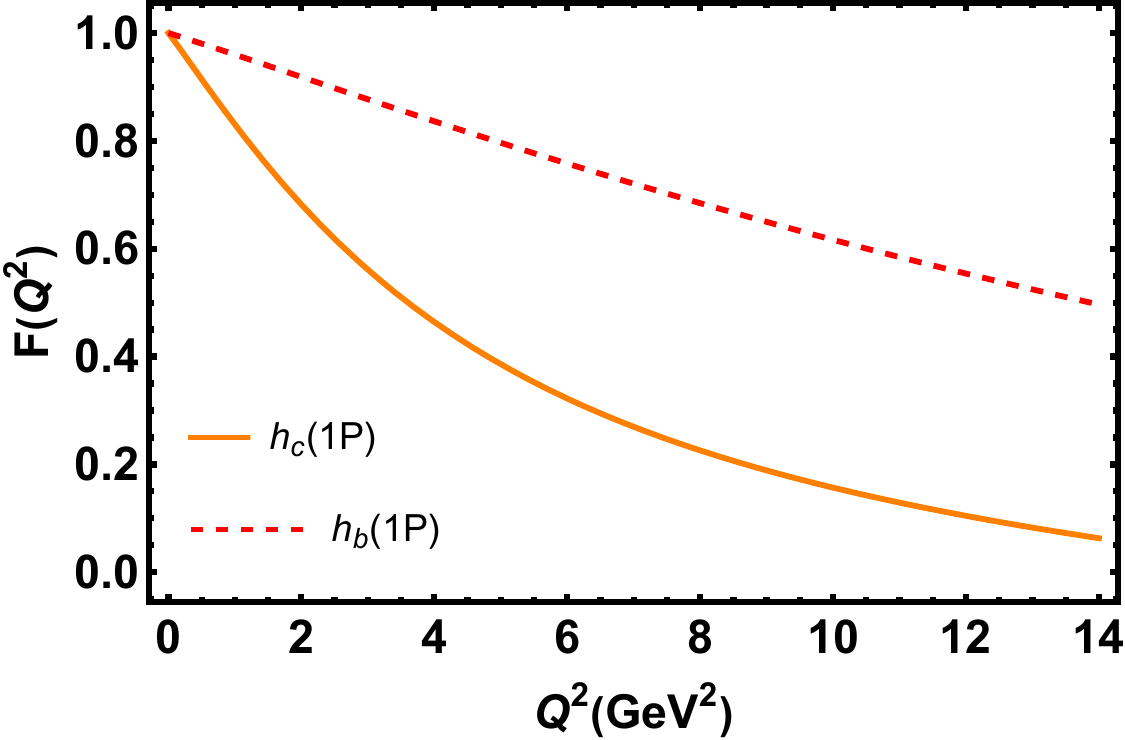}
		\caption{} \label{(d1)}
	\end{subfigure}
	\caption{Form factors of $\eta_{c}$ (orange solid line) and $\eta_{b}$ (red dashed line) for $1S$, $2S$, $3S$ and $1P$ state. The $\eta_{c}$ $1S$ state has been compared to the Lattice $B1$ and Lattice $C1$ simulation data \cite{Li:2020gau}. \label{fig:7}}
\end{figure}

In Fig. \ref{fig:7}, we have compared the EMFFs of $\eta_{c}$ and $\eta_{b}$ for the various excited states. As the form factors of heavy quarkonia derived from the combined quark and anti-quark contributions disappear, we have present solely the quark portion of the contribution. We have compared our results for the $1S$ and $1P$ state $\eta_{c}$ meson with the lattice $B1$ and $C1$ simulation data \cite{Li:2020gau} and our results are in good agreement with theirs. We find that the form factors for the $\eta_{b}$ meson are in general steeper than those for the $\eta_{c}$ meson and the curves saturate at much larger $Q^{2}$ values. Since we have identical quark and anti-quark masses in our quarkonia system, the contributions from both the quark and anti-quark are almost equally responsible for the distribution in the $Q^{2}$ region.

In Table \ref{charge rad}, we have presented the calculated charge radii $\langle r^{2} \rangle$ of the excited states heavy quarkonia. We have compared our calculations with the theoretical results of the lattice simulations \cite{Li:2020gau} and BLFQ \cite{Li:2015zda} since we lack any experimental data regarding this topic. We find that our results agree well with the lattice calculations. As per our understanding of charge radii, the heavy quark is seated near to the meson's core while the light quark is traveling and the light quark's mobility in a heavy-light system primarily determines the corresponding meson's charge radius. For our case of the heavy-heavy system, both the heavy quarks are situated close to the center of the meson with little to no mobility and hence we obtain very low values of the charge radii. This is evident while comparing the outcomes of LFQM from Ref. \cite{Hwang:2001th}.

\begin{table}[]
	\centering
	\begin{tabular}{|   c   |   c   |   c   |}
		\hline \hline & $\langle r^{2} \rangle_{\eta_{c}(1 S)}$  & $\langle r^{2} \rangle_{\eta_{b}(1 S)}$  \\
		\hline Our results &  0.041& 0.010  \\
		BLFQ \cite{Li:2015zda} & 0.038(5) &  0.0146(8) \\
		DSE \cite{Maris:2006ea} &  0.048(4) &  \\
		Lattice, B1 \cite{Li:2020gau} & 0.052(4) &  \\
		\multirow[t]{2}{*}{ Lattice, C1 \cite{Li:2020gau} } & 0.044(4) &  \\
		\hline
		\hline & $\langle r^{2} \rangle_{\eta_{c}(2 S)}$ & $\langle r^{2} \rangle_{\eta_{b}(2 S)}$ \\
		\hline Our results & 0.089 & 0.020  \\
		\multirow[t]{2}{*}{BLFQ \cite{Li:2015zda}} & 0.1488(5) & 0.0510(8) \\
		\hline
		\hline & $\langle r^{2} \rangle_{\eta_{c}(3 S)}$ & $\langle r^{2} \rangle_{\eta_{b}(3 S)}$ \\
		\hline \multirow[t]{2}{*}{Our results} & 0.107 & 0.037 \\
		\hline
		\hline & $\langle r^{2} \rangle_{h_{c}(1 P)}$ & $\langle r^{2} \rangle_{h_{b}(1 P)}$ \\
		\hline Our results & 0.126 & 0.056 \\
		\hline \hline 
	\end{tabular}
	\caption{Charge radii (in fm$^{2}$) of heavy quarkonia in comparison to other theoretical models.}
	\label{charge rad}
\end{table}

\subsection{Parton Distribution Functions (PDFs)} 
PDFs give us 1-D picture of the hadron and describe the likelihood of  finding a parton carrying the parent hadron's longitudinal momentum fraction ($x$).   Using the LF approach, the PDFs of a hadron state can be determined by integrating the transverse momentum of the wave function overlaps. The unpolarized quark's PDF for spin-0 mesons is defined as

\begin{equation}
      f_{1}(x)= \int \frac{\mathrm{d}^{2} \mathbf{k}_{\perp}}{(2 \pi)^{3}}\left|\phi_{nS(nP)}\left(x, \mathbf{k}_{\perp}\right)\right|^{2}.
      \end{equation}
We normalize this PDF to unity in our model as
\begin{equation}
    \int_{0}^{1} f_{1}(x) dx = 1.
\end{equation}
For the spin-1 mesons, there are four T-even distributions for the twist-2 case which are  $f_{1}(x)$, $f_{1 LL}(x)$, $h_{1}(x)$ and $g_{1} (x)$, depending on the relative polarization of both the quark and its respective meson. We can simply integrate the T-even TMDs over the transverse momentum $\mathbf{k}_{\perp}$ to obtain the analytical form for the spin-1 PDFs. For the present work, we have only considered the  unpolarized PDF $f_{1}(x)$ which describes the distribution of the unpolarized quark in the unpolarized spin-1 meson.

For the heavy vector mesons, the overlap form of the LFWFs with respect to the LF helicity amplitudes with quark-antiquark helicities and their polarizations can be written as \cite{Puhan:2023hio}

\begin{figure}[htbp]
	\centering
	\begin{subfigure}[b]{0.49\textwidth}
	{\includegraphics[width=\linewidth]{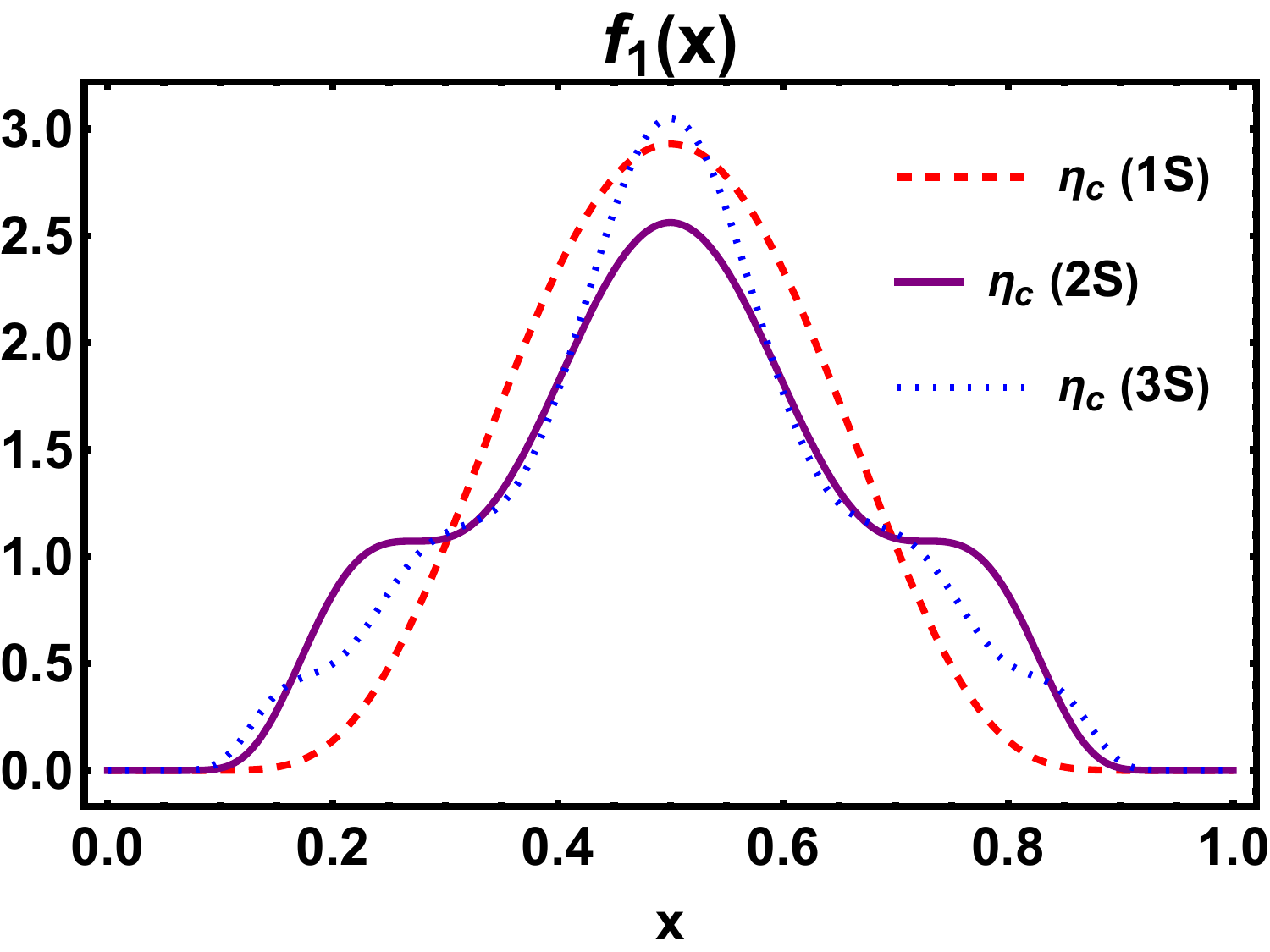}}
		\caption{} \label{(PDF1)}
	\end{subfigure}
    \begin{subfigure}[b]{0.49\textwidth}
	{\includegraphics[width=\linewidth]{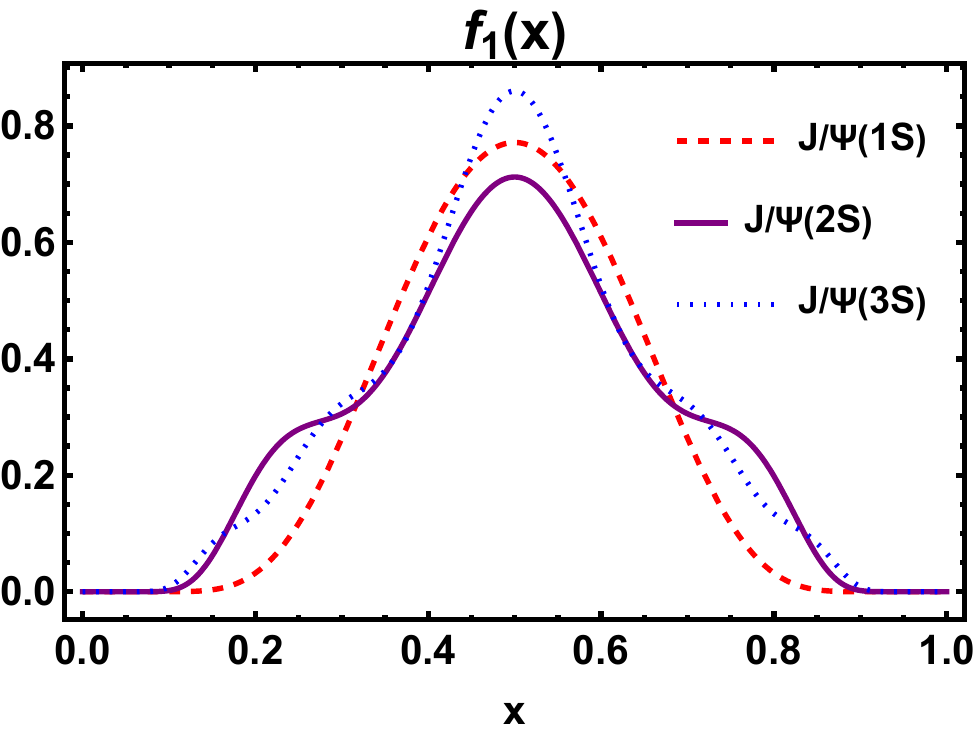}}
		\caption{} \label{(PDF2)}
    \end{subfigure}
	\begin{subfigure}[b]{0.49\textwidth}
	{\includegraphics[width=\linewidth]{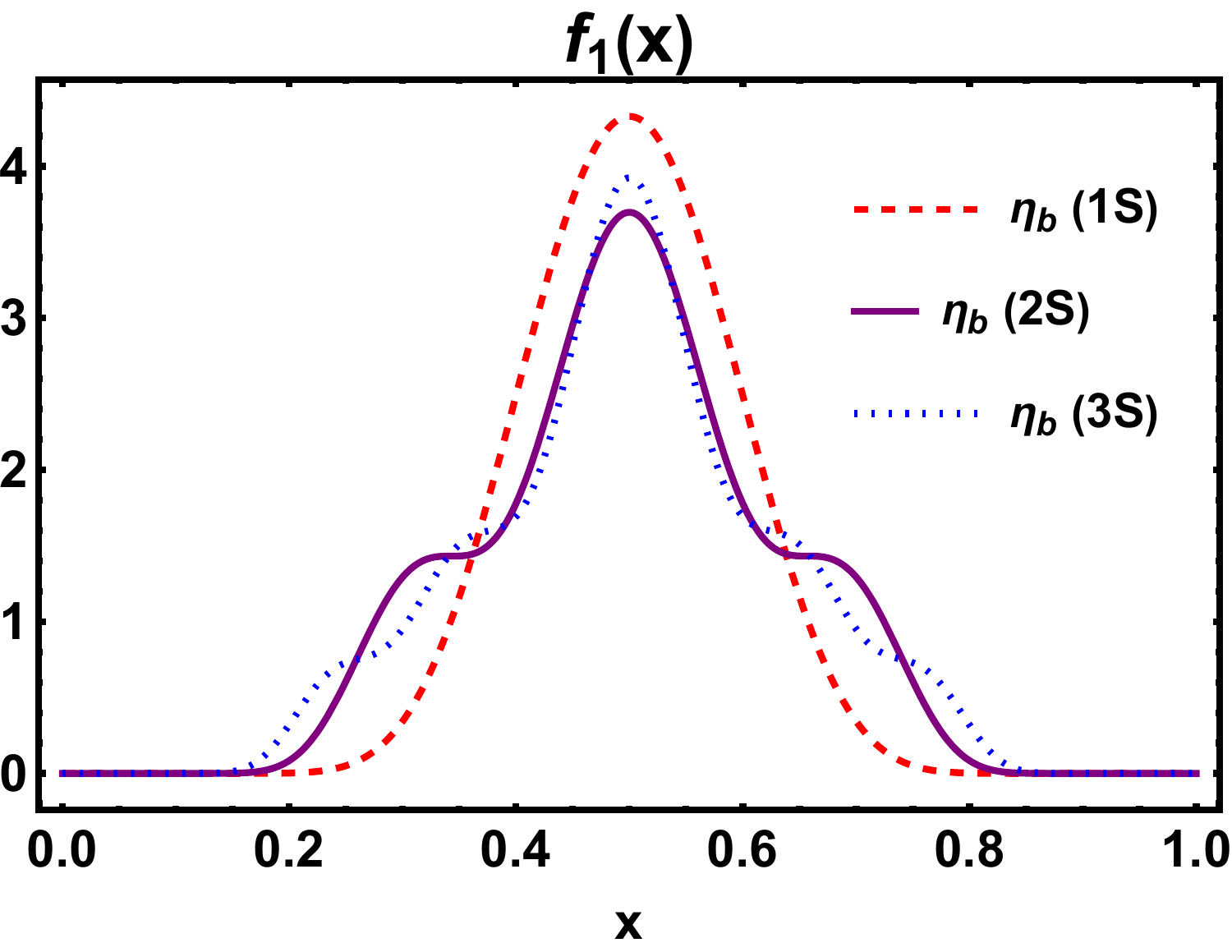}}
		\caption{} \label{(PDF3)}
	\end{subfigure}
    \begin{subfigure}[b]{0.49\textwidth}
	{\includegraphics[width=\linewidth]{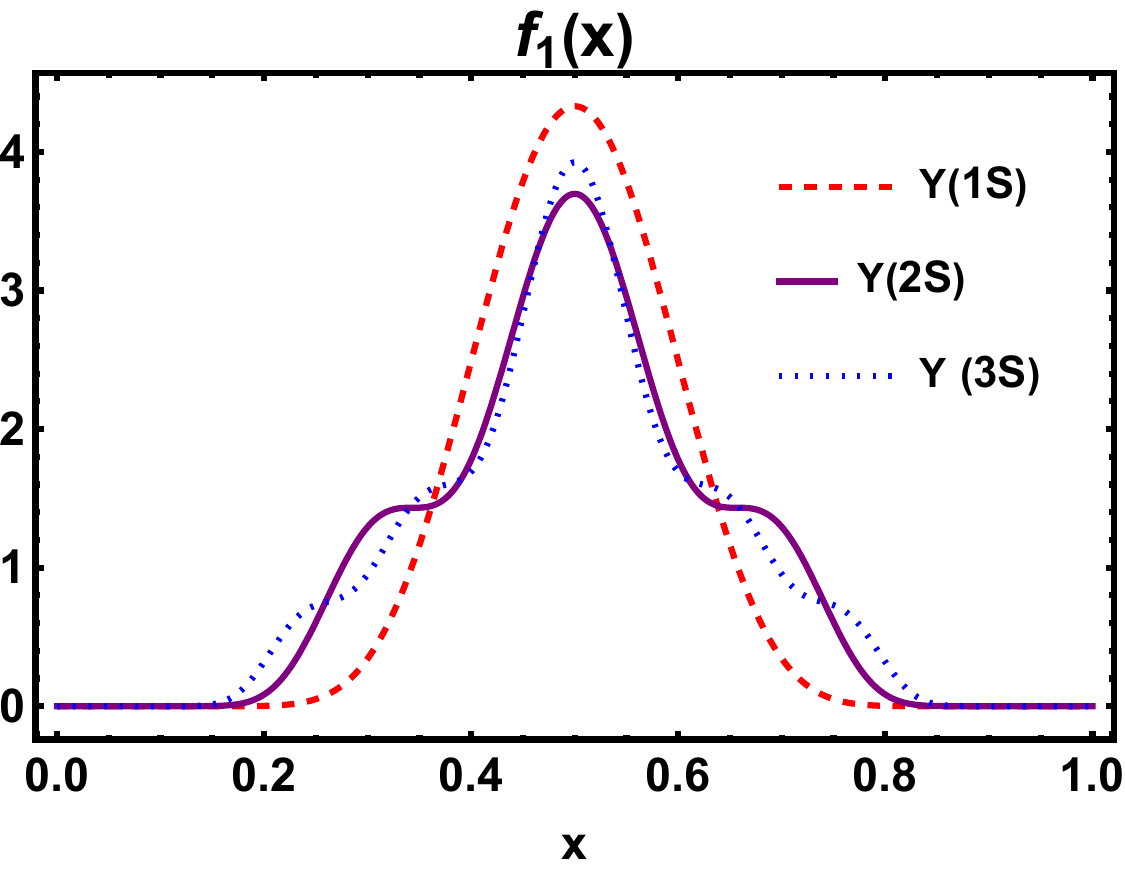}}
		\caption{} \label{(PDF4)}
	\end{subfigure}
	\qquad
	\begin{subfigure}[b]{0.49\textwidth}
	{\includegraphics[width=\linewidth]{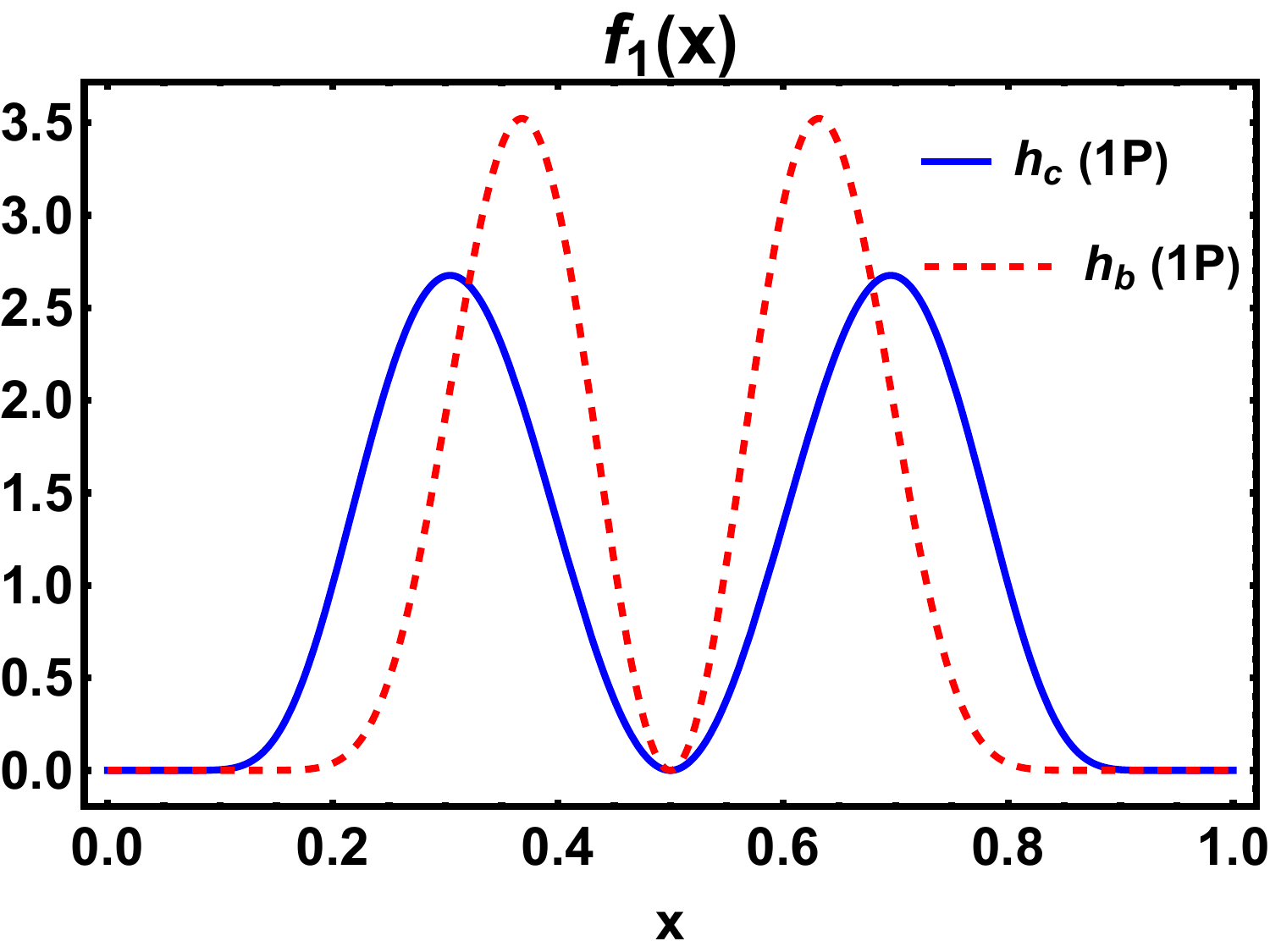}}
		\caption{} \label{(PDF5)}
	\end{subfigure}
    \begin{subfigure}[b]{0.49\textwidth}
	{\includegraphics[width=\linewidth]{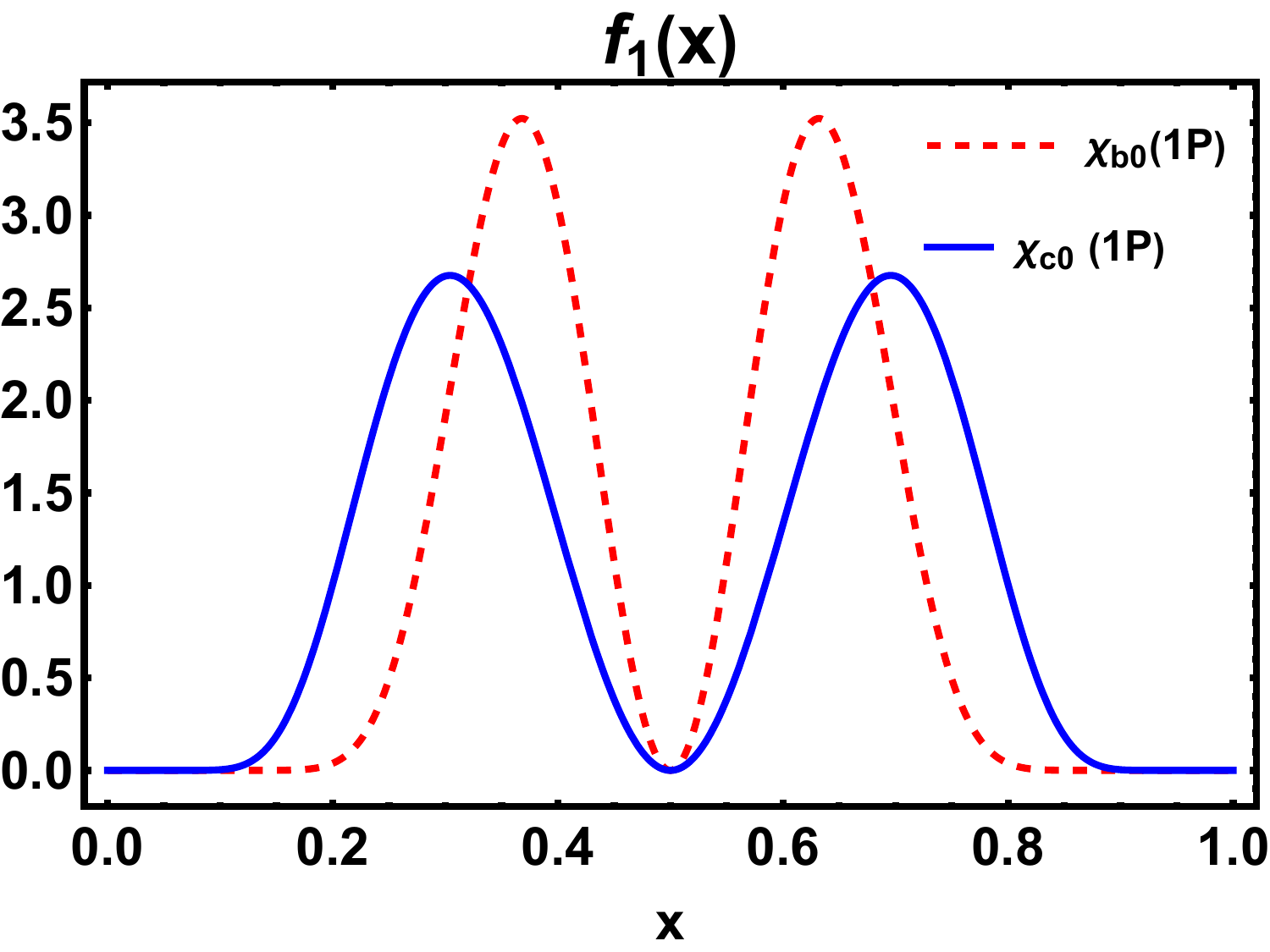}}
		\caption{} \label{(PDF6)}
	\end{subfigure}
	\caption{$f_{1} (x)$ PDFs plotted as a function of $x$ for the radially excited states of the  $\eta_{c}$,  $J/\Psi$,  $\eta_{b}$,  $\Upsilon$ in (a), (b), (c) and (d) respectively. The comparison of the orbitally excited state for $h_{c}$and $h_{b}$ has been shown in (e) whereas comparison of  $\chi_{c0}$ and $\chi_{b0}$ has been shown in (f).}\label{Tom23}
\end{figure}

\begin{equation}
\mathrm{A}^{\Lambda^{\prime},\Lambda}_{\lambda_q^{\prime}, \lambda_q}\left(x, \mathbf{k}_{\perp}\right)=\frac{1}{(2 \pi)^3} \sum_{\lambda_{\bar{q}}} \psi_{\lambda_q^{\prime},\lambda_{\bar q}}^{J J_{z}*}\left(x, \mathbf{k}_{\perp}\right) \psi_{\lambda_q,\lambda_{\bar q}}^{J J_{z}}\left(x, \mathbf{k}_{\perp}\right),
\end{equation}
where, $\Lambda$ is the spin projection of the spin-1 meson.
For the unpolarized T-even TMD, the explicit overlap form in terms of helicity amplitudes is expressed as \cite{Puhan:2023hio}

\begin{equation}
    f_1\left(x, \mathbf{k}_{\perp}^2\right)= \frac{1}{6}\left(\mathrm{A}^{0,0}_{+,+}+\mathrm{A}^{0,0}_{-,-}\right. 
\left.+\mathrm{A}^{+,+}_{+,+}+\mathrm{A}^{+,+}_{-,-}+\mathrm{A}^{-,-}_{+,+}+\mathrm{A}^{-,-}_{-,-}\right).
\end{equation}
Therefore, using the above we can obtain the analytical form for the unpolarized $f_{1} (x)$ PDF from its corresponding $f_1\left(x, \mathbf{k}_{\perp}^2\right)$ TMD as \cite{Puhan:2023hio}

\begin{align}
f_{1} (x) &= \int \mathbf{d^{2} k_{\perp}} \frac{1}{3 (2 \pi)^{3}} 
\Bigg[
\frac{1}{2} \left( 3 (m ( M + 2m))^{2} + (1 - 2 x)^{2} M^{2} \mathbf{k}_{\perp}^{2} \right) \\
&\qquad + 4 \mathbf{k}_{\perp}^{2} (m (M + 2m) + \mathbf{k}_{\perp}^{2}) \nonumber \\
&\qquad + \mathbf{k}_{\perp}^{2} \left( 2m (M + m) + (1 - 2x + x^{2}) M^{2} \right) 
\Bigg]
\frac{\left| \psi_{\lambda_q^{\prime},\lambda_{\bar q}}^{J J_{z}}\left(x, \mathbf{k}_{\perp}\right) \right|^{2}}{w^{2}} \nonumber,
\end{align}
where 
\begin{equation}
	w = (2m + M) \sqrt{\mathbf{k}_{\perp}^2 + m^{2}}.
\end{equation}

\begin{figure}[!htb]
	\centering
	{\includegraphics[width=.49\linewidth]{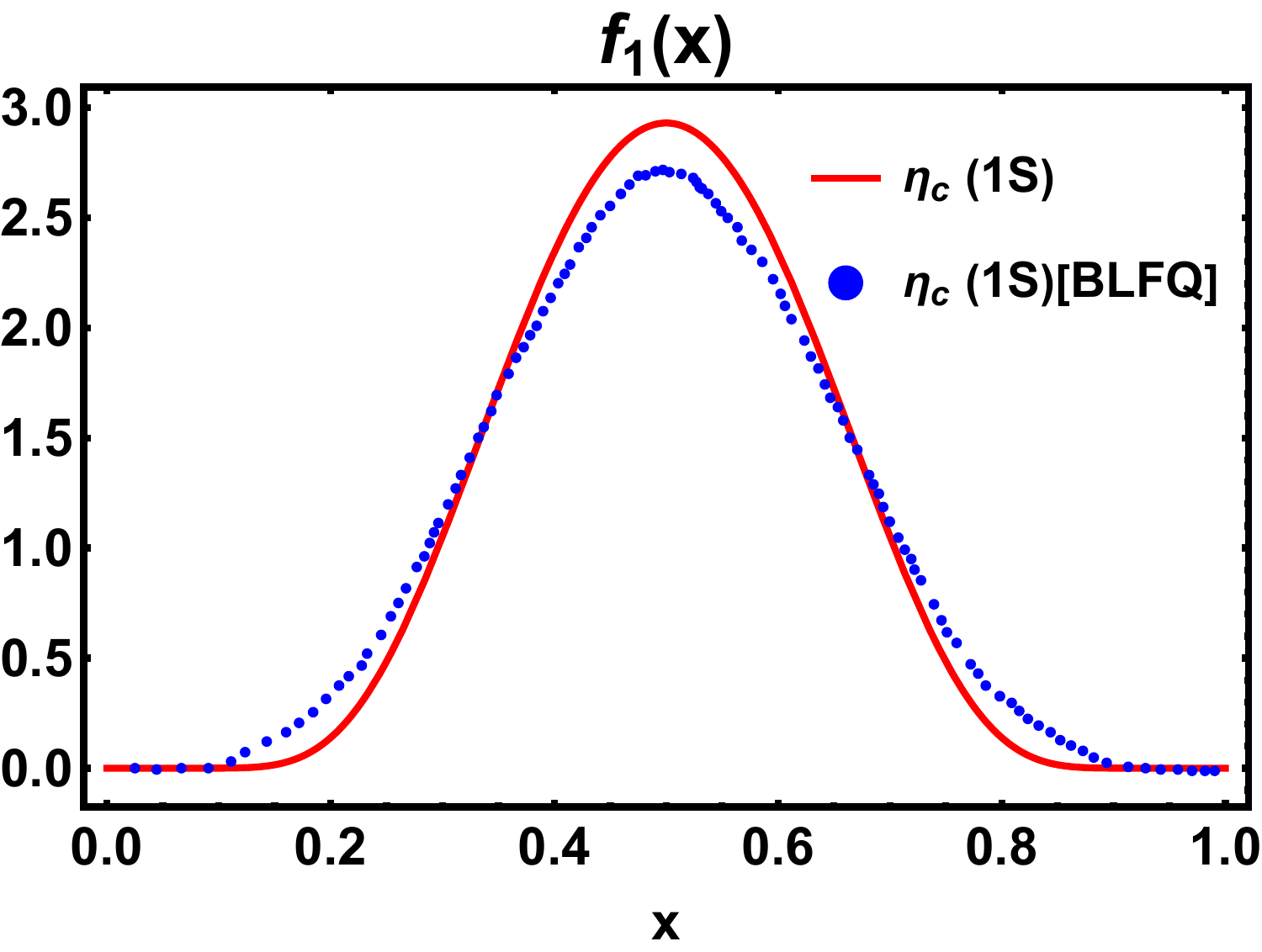}}
	{\includegraphics[width=.49\linewidth]{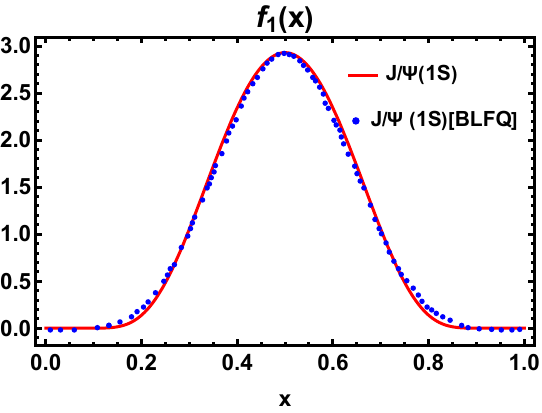}}
    {\includegraphics[width=.49\linewidth]{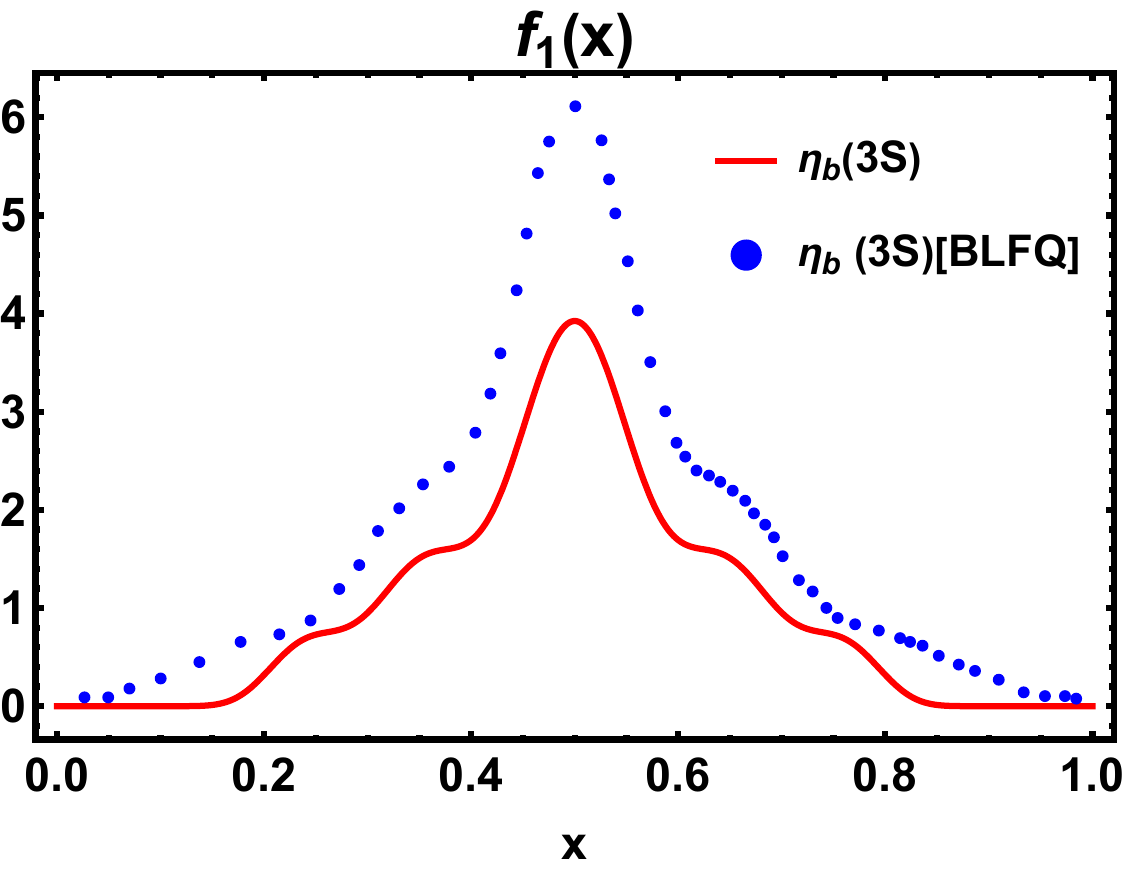}}
    {\includegraphics[width=.49\linewidth]{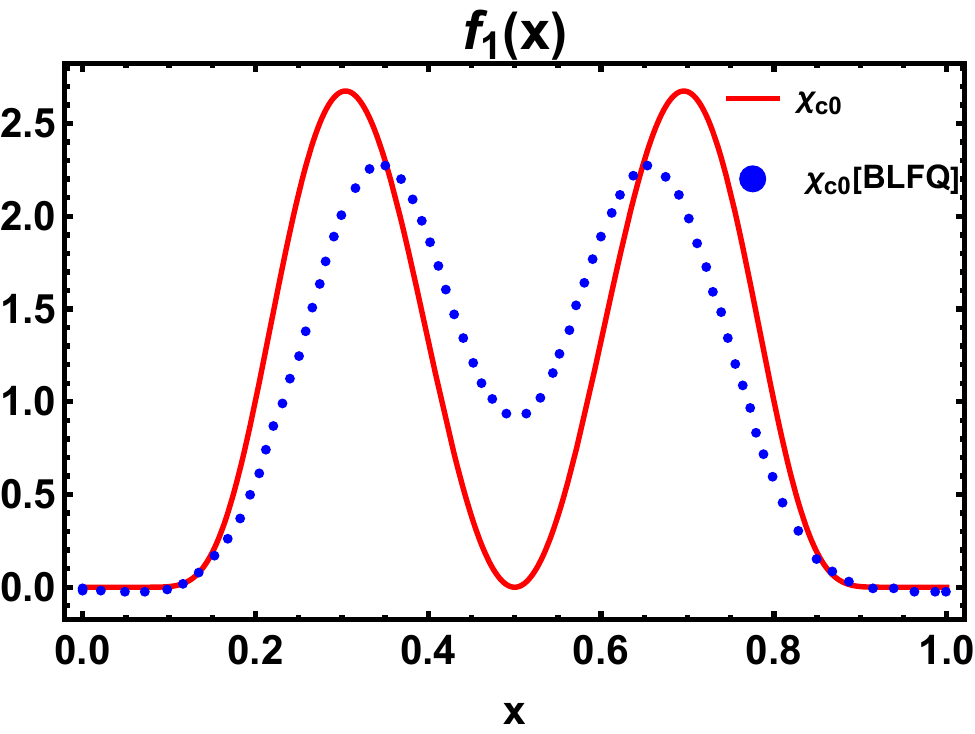}}
	\caption{$f_{1} (x)$ PDFs for the $1S$ state $\eta_{c}$ meson and $3S$ state $\eta_{b}$ meson (left panel), along with $1S$ state $J/\psi$ and $1P$ state $\chi_{c0}$ (right panel) in comparison with the BLFQ data \cite{Lan:2019img,Li:2017mlw}.}\label{Tom235677}
\end{figure}

In Fig. \ref{Tom23}, we have  presented the unpolarized PDFs as a function of longitudinal momentum fraction $x$ for the various excited states of the spin-0 and spin-1 mesons. The plots are symmetric about $x = 0.5$ which is symbolic to their similar quark and anti-quark masses. Similar to the DAs, we observe  here that the bottomonia ($b \bar b$) PDFs have a narrower waveform along with a higher amplitude peak as compared to the charmonia ($c \bar c$) PDFs. As we go to higher excited states, the PDFs tend to have ridges along their curves which is due to their nodal wave structure. We also notice that the $1P$ state has a double amplitude which we believe arises due to the $k_{z}$ factor present in the orbital wave function that we have considered. We conclude that when the quarks reach higher orbital states, they tend to carry a very high percentage of the momentum fraction $x$ inside the parent hadron.

PDFs for the charmonia ($c \bar c$) and bottomonia ($b \bar b$) have been studied in detail in other theoretical formulations such as in BLFQ. We have compared our results for the $1S$ state $\eta_{c}$ meson, $3S$ state of $\eta_{b}$ meson, $1S$ state of the $J/\psi$ meson and the $1P$ state of $\chi_{c0}$ to the ones calculated in the BLFQ method \cite{Lan:2019img,Li:2017mlw} in Fig. \ref{Tom235677}. Our results seem to be in fairly good agreement in most of the cases except in the $3S$ and $1P$ states which vary mostly in their amplitudes.
This difference is again due to different model assumptions and input parameters.

\section{Summary} \label{sum}

In the present work, we have studied the ground state $1S$, radially excited states $2S$ and $3S$ along with the orbitally excited state $1P$ for the heavy quarkonia in both spin-0 and spin-1 mesons in the LFQM. The results obtained for the decay constants are in good agreement with the experimental data and lattice simulations for the $1S$ state but the similarities deviate as we go to higher excited states. This contradicts the empirical hierarchy that the decay constants should decrease with the increase in the excitation of the hadrons. Introducing mixing effects to the $1S$ and $2S$ states has already been studied and mixing of higher radially excited states in the present study further gives required results for the decay constants. For the case of DAs, our results are in good agreement with the lattice simulations. We note that for the $1S$ state, there is no visible difference between the pseudo-scalar and vector DAs but the differences get more pronounced and prominent as we go to the higher excited states. We have also observed that the DAs for the bottomonium systems have a narrower shaped wave form as compared to the charmonium systems. In addition, the 3-D plots presented for the $\eta_{c}$ meson for all the excited states give a better visualization of the wave functions we employ in this work. The PDFs considered in this work have given us a more precise understanding of the internal structure of the hadron. No nodes are observed for the  $1S$ and $1P$ states where as nodes exist for the $2S$ and $3S$ states. In this case also, the bottomonium systems have a narrower shaped wave form as compared to the charmonium systems. The results of EMFFs and charge radii computed for $\eta_{c}$ and $\eta_{b}$ mesons for their excited states are in good agreement with the available lattice simulation data. We have mainly addressed the heavy-heavy meson sector in the framework of LFQM for this work. Nevertheless, a collective investigation of the heavy and light meson sectors is equally crucial to examine how closely physical quantities depend on quark masses. Accurate measurements of the physical characteristics of heavy mesons and findings of yet unidentified heavy meson states in future will be crucial to evaluate phenomenological notions about the heavy meson structure.
\section{Acknowledgement}
NK and HD would like to acknowledge the research grant received from the Science and Engineering Research Board, Government of India under Teachers Associateship Research Excellence Award, Grant No. TAR/2021/000157.

\end{document}